\documentclass[twocolumn,twocolappendix,tighten]{aastex701}
\usepackage{amsmath, amssymb, mathtools, bm}

\usepackage{graphicx}
\usepackage[caption=false]{subfig} 

\usepackage{booktabs, tabularx}
\usepackage[flushleft]{threeparttable}

\usepackage{paralist}

\usepackage{color}
\usepackage{comment}

\usepackage{float}
\usepackage{changepage}
\usepackage{perpage}
\MakePerPage{footnote}

\begin{document}


\title{Impact of Cosmic Ray Distribution on the Growth and Saturation of Bell Instability}

\author[0009-0002-7852-1247]{Saikat Das}
\affiliation{Joint Astronomy Programme, Department of Physics, Indian Institute of Science, Bangalore 560012, India}
\email[show]{saikatdas@iisc.ac.in}

\author[0000-0002-1030-8012]{Siddhartha Gupta}
\affiliation{Department of Astrophysical Sciences, Princeton University, 4 Ivy Ln., Princeton, NJ 08544, USA}
\email{gsiddhartha@princeton.edu}

\author[0000-0003-2635-4643]{Prateek Sharma}
\affiliation{Joint Astronomy Programme, Department of Physics, Indian Institute of Science, Bangalore 560012, India}
\email{prateek@iisc.ac.in}



\begin{abstract}

Cosmic rays (CRs) streaming in weakly magnetized plasmas can drive large-amplitude magnetic fluctuations via nonresonant streaming instability (NRSI), or Bell instability. 
Using one-dimensional kinetic simulations, we investigate how mono-energetic and power-law CR momentum distributions influence the growth and saturation of NRSI. 
The linear growth is governed solely by the CR current and is largely insensitive to the CR distribution.
However, the saturation depends strongly on the CR distribution and is achieved through CR isotropization, which quenches the driving current.
Mono-energetic CRs effectively amplify the magnetic field and isotropize.
For power-law distributions, the lowest-energy CRs dominate current relaxation and magnetic growth, while the highest-energy CRs remain weakly scattered, limiting their contribution to saturation.
In the absence of low-energy CRs, high-energy particles amplify magnetic fields effectively and isotropize.
Accounting for these effects, we provide a modified saturation prescription valid for both relativistic and nonrelativistic CRs.
We propose a layered CR-confinement scenario upstream of astrophysical shocks, relevant to particle acceleration to high energies.

\end{abstract}



\section{Introduction} \label{sec:intro}

A central challenge in high-energy astrophysics is to understand the acceleration, propagation, and confinement of cosmic rays (CRs).
The diffusive shock acceleration (DSA) at a strong shock provides a robust framework for particle acceleration where the maximum energy of CRs relies critically on their confinement \citep{axford1977dsa,bell1978dsa,blandford1978particle,krynskii1977dsa}. 
Without sufficient electromagnetic turbulence, CRs escape too quickly due to their large Larmor radii, limiting their maximum energy \citep{hillas1984origin}.  
Identifying the physical origins of such turbulence and its dependence on plasma parameters is essential to complete our understanding of CR acceleration and its backreaction on the background plasma.

Nonthermal X-ray and radio observations of supernova remnants (SNRs) reveal magnetic fields of several hundred microgauss in their vicinity, far exceeding those typically found in the interstellar medium (ISM) \citep{berezhko2003, vink_2003, volk2005magnetic}.
This suggests that CR-accelerating sites are influenced by local magnetic field amplification, either due to preexisting turbulence or self-generated instabilities driven by the CRs themselves \citep{eriksen_2011,matsuda_2020}.
The latter, also known as CR streaming instabilities, are expected to arise naturally due to the relative drift between CRs and the thermal plasma upstream of SNR shocks \citep{mckenzie1982non,bykov2013microphysics,zweibel2013microphysics}.

The long-known CR resonant streaming instability is responsible for CR scattering and magnetic field amplification in the interstellar and intracluster medium \citep{lerche1967unstable,kulsrud1969effect,wentzel1969propagation}. However, it is generally limited to modest amplification levels, with $\delta B/B_0 \lesssim 1$ \citep{bai_2019,holcomb_2019}.
A major advance came with the finding of CR–driven {\it nonresonant instabilities}, capable of amplifying magnetic fields well beyond background levels \citep{bell2001,bell2004}.
Upstream plasma in strong shocks, such as SNRs, provides ideal conditions for significant NRSI-driven magnetic field amplification. 

Although both resonant and nonresonant streaming instabilities (hereafter, RSI and NRSI, respectively) generate electromagnetic turbulence,  their mechanisms differ significantly.
The RSI occurs when the transverse fluctuations grow by resonantly interacting with the gyromotion of the CRs. 
In contrast, NRSI is driven by a strong super-Alfv\'{e}nic CR current, which grows transverse magnetic fluctuations via the $\bm{J}\times \bm{B}$ force.

The NRSI (Bell instability) has been studied extensively using analytical approaches, and kinetic, hybrid, and magnetohydrodynamic (MHD) simulations \citep{niemiec2008production,zirakashvili2008diffusive,bret2009weibel,riquelme2009nonlinear,amato2009,gargate2010nonlinear,reville2013universal,matthews2017amplification,weidl2019background}.
These studies explored key aspects, including magnetic field saturation \citep{bell2004, gupta_2021, zacharegkas2024modeling}, effects of background plasma temperature \citep{zweibel2010environments, marret2021growth}, and the differences between proton and lepton-driven NRSIs \citep{gupta_2021}, using {\it mono-energetic} CRs. 
However, a systematic study of how NRSI depends on the shape of the CR distribution is still lacking  \citep{Haggerty:2019anu}, particularly the {\it power-law} momentum distribution $f(p)\propto p^{-4}$ expected from DSA.

The present work focuses on the impact of different CR distributions during the linear growth and nonlinear saturation phases of NRSI, which has not been explored before.
We use first-principles kinetic plasma simulations to address the following questions.
\begin{compactitem}
\setlength\itemsep{0.07em}
    \item Does the linear regime of the instability depend on the distribution of CRs?
    \item How does the saturated magnetic field depend on changes to the minimum momentum ($p_{\rm min}$) or the maximum momentum ($p_{\rm max}$) of the power-law CR distribution?
    \item Does the saturation differ between a mono-energetic and a power-law CR distribution with similar bulk properties?
\end{compactitem}
The paper is organized as follows. 
Section \ref{sec:theory} goes through the analytical approaches for NRSI. 
Section \ref{sec:numerical_setup} outlines our numerical setup of kinetic simulations, and Section \ref{sec:results} shows the results. 
We discuss the astrophysical implications of our findings in Section \ref{sec:discussion}, before concluding in Section \ref{sec:conclusion}.

\section{Analytic Theory} \label{sec:theory}
We analytically study the linear growth (Section \ref{sec:analytical_linear_growth}) and saturation (Section \ref{sec:analytical_saturation}) of NRSI, comparing them for mono-energetic and power-law CR distributions.

\subsection{Linear Growth} \label{sec:analytical_linear_growth}

We derive the linear dispersion relations for NRSI following standard methods. 
We review the fluid approach (Section \ref{sec:analytical_fluid_approach}) and examine the impact of CR distribution functions with the kinetic approach (Section \ref{sec:analytic_kinetic_approach}).
We show that the fastest-growing NRSI modes are largely insensitive to the CR distributions (Section \ref{sec:analytical_growth_comparison}).

\subsubsection{Fluid approach} \label{sec:analytical_fluid_approach}

We consider a plasma with electrons and ions having number densities $n_e$ and $n_i$, respectively, and CRs with number density $n_{\text{cr}}$, much smaller than $n_e$, $n_i$.
The charge-neutrality of plasma requires
\begin{equation} \label{eq:charge_nutral}
    n_{\text{cr}} + n_i = n_e.
\end{equation}
The CRs drift along the background magnetic field ($B_0 \bm{\hat{x}}$) with a drift speed of $v_d$.
To maintain current neutrality in the plasma, the background electrons quickly develop a return current \citep{amato2009,gupta+24a}, with a drift speed 
\begin{equation} \label{eq:current_nutral}
    v_e = \frac{n_{\text{cr}}}{n_e} v_d.
\end{equation}

Treating CRs as unperturbed during the linear phase, the NRSI growth rate is
\begin{equation} \label{eq:growth_rate_fluid}
    \gamma_{\rm fluid} (k) = v_{A0} k \left[ \frac{k_{\text{u}}}{k} - 1 \right]^{1/2}
\end{equation}
\citep{gupta_2021}. Here, $v_{A0} = B_0 / \left( 4 \pi m_i n_0 \right)^{1/2}$ is the Alfv\'en speed, $n_0 = n_e \approx n_i$ is background plasma density, $k$ is parallel wavenumber, and 
\begin{equation} \label{eq:k_u_fluid}
    k_{\text{u}} = \frac{v_e}{v_{A0}^2} \omega_{ci} = \frac{n_{\text{cr}}}{n_e} \frac{v_d}{v_{A0}^2} \omega_{ci},
\end{equation}
with $\omega_{ci}=eB_0/m_i c$, the cyclotron frequency of background ion.
Modes with $k<k_{\text{u}}$ are unstable. 

From Equation \ref{eq:growth_rate_fluid}, the fastest-growing mode $k_{\rm fast}$ and its growth rate $\gamma_{\rm fast}$ are evaluated as
\begin{equation} \label{eq:k_fast_fluid}
    k_{\rm fast} d_i = \frac{1}{2} \frac{n_{\rm cr}}{n_e} \frac{v_d}{v_{A0}} = \frac{\gamma_{\rm fast}}{\omega_{ci}},
\end{equation}
where $d_i=v_{A0}/ \omega_{ci}$ is the skin depth of background ions.
When $k_{\rm fast} \left( = k_{\rm u}/2\right)$ is much larger than the inverse of the Larmor radius of even the lowest-energy CR particle ($k_{\rm fast} \gg 1/r_{L,0}$), the instability becomes non-resonant.

\subsubsection{Kinetic approach} \label{sec:analytic_kinetic_approach}

We define the CR distribution functions as 
\begin{equation} \label{eq:distribution_functions}
    f'_{\text{cr}} \left( p,\mu \right) = \frac{n_{\text{cr}}}{2 \pi} g' \left( p,\mu \right),
\end{equation}
where $g'$ is normalized such that
\begin{equation} \label{eq:normalization_condition}
  \int_0^\infty dp \int_{-1}^1 d\mu \left[ 2\pi p^2 f'_{\text{cr}} \left( p,\mu \right) \right] = n_{\text{cr}},
\end{equation}
and $\mu=p_x/p$ is the pitch-angle, $p_x$ being the $x$-component (background field direction) of momentum $p$.
The apostrophe ($'$) denotes the CR-frame, the frame in which the CR distribution function is conveniently specified and not necessarily the frame in which the CR bulk velocity vanishes. 

We consider two CR momentum distributions: mono-energetic (ME) and power-law (PL).
The ME CRs are defined with
\begin{equation} \label{eq:distribution_ME}
    g'_{\text{ME}}(p,\mu)= \left[ \frac{1}{p^2} \delta \left( p-p'_0 \right) \right] \left[ \frac{\Theta \left( \mu -\mu'_{\min} \right)}{1-\mu'_{\min}} \right],
\end{equation}
where the delta ($\delta$) and Heaviside ($\Theta$) functions are used.
The PL CRs are specified by
\begin{equation} \label{eq:distribution_PL}
\begin{aligned}
    g'_{\text{PL}}(p,\mu)&= \left[ \frac{p^{-4}}{\left( p'^{-1}_{\text{min}} - p'^{-1}_{\text{max}} \right)} \right] \left[ \frac{\Theta \left( \mu -\mu'_{\min} \right)}{1-\mu'_{\min}} \right],
\end{aligned}
\end{equation} 
with momentum limits $p'_{\text{min}} \leq p' \leq p'_{\text{max}}$.

The parameter $\mu'_{\min}$ determines the pitch-angle distributions of CRs: $\mu'_{\min}=-1$ corresponds to {\it isotropic} or spherical distributions, while $\mu'_{\min}=0$ yields a forward {\it cone} or hemispherical distribution (i.e. all the CRs have positive $p_x$). 
The isotropic CRs require additional boost velocity $v_b \bm{\hat{x}}$ to drift along $B_0 \bm{\hat{x}}$ relative to the background plasma. 
The cone CRs naturally have a net drift and do not require additional boost to generate a current.
The cone distribution qualitatively represents CRs streaming in the ISM with limited scattering.

Using these CR distributions, we derive NRSI growth rates by linearizing the Vlasov equation, following the approach in \citet{amato2009}.
Appendix \ref{sec:derivation_kinetic_dispersion} contains a brief summary of the dispersion relations.

\subsubsection{Linear growth rates for different CR distributions} \label{sec:analytical_growth_comparison}

\begin{figure}
    \centering
    \includegraphics[width=0.47\textwidth]{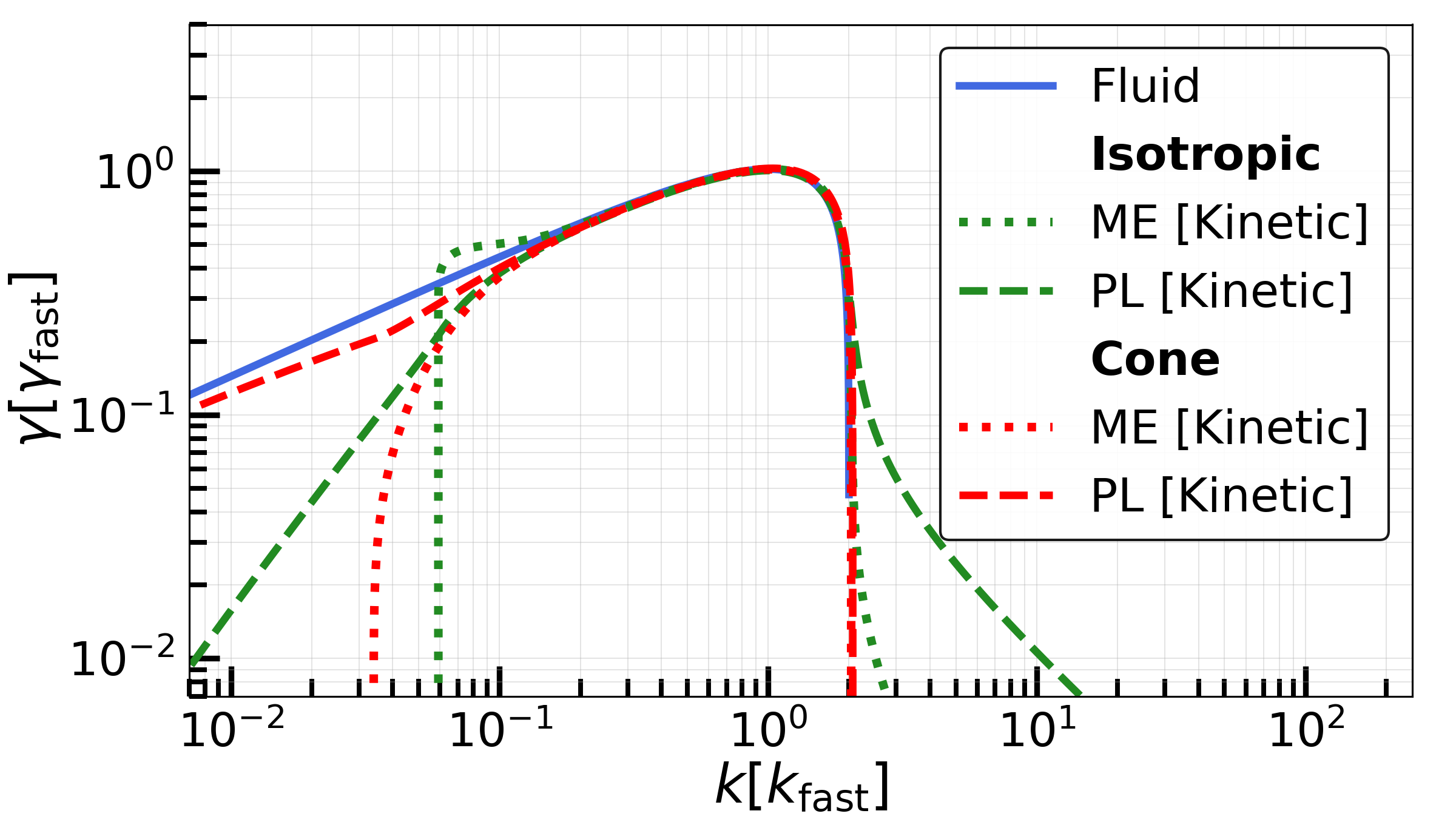}
    \caption{Analytic growth rates of NRSI derived for mono-energetic (ME, dotted) and power-law (PL, dashed) momentum distributions of CRs. 
    The blue curve represents the growth rate obtained from the fluid approach (Equation \ref{eq:growth_rate_fluid}), while all other curves are obtained from the kinetic approach (Equations \ref{eq:dispersion_iso} and \ref{eq:dispersion_cone}).
    The maximum growth rates and the corresponding wavenumbers remain almost the same; they depend on the bulk CR current. } 
    \label{fig:dispersion_analytical}
\end{figure}

We compare the NRSI growth rates obtained from the kinetic approach for different CR distributions with the fluid dispersion, which, by definition, is agnostic to CR distributions.
Figure \ref{fig:dispersion_analytical} shows these growth rates, where we have used the parameters $v_{A0}=0.02c$, $n_{\text{cr}}/n_e=0.01$, $p'_0=p'_{\text{min}} \approx 5 m_i c$, and $p'_{\text{max}} \approx 400 m_i c$. 
For isotropic and cone CR distributions, the drift speeds are $v_d \approx 0.3c$ and $v_d \approx 0.5c$, respectively. 
We normalize the wavenumbers $k$ and growth rates $\gamma$ with $k_{\rm fast}$ and $\gamma_{\rm fast}$ (as Equation \ref{eq:k_fast_fluid}).
Note that we use the same parameters later in our simulations (see Section \ref{sec:numerical_setup}).

For all cases in Figure \ref{fig:dispersion_analytical}, the maximum growth rate and its wavenumber closely match the fluid approach, implying that the fastest-growing NRSI modes depend only on the net CR current.
However, $\gamma$ for mono-energetic CRs drops sharply below $k \lesssim 1/r'_{L,0}$ ($r'_{L,0}=p'_0c/eB_0$ is the Larmor radius), where CRs become resonant, and their kinetic response suppresses NRSI, unlike the fluid derivation.
In contrast, for power-law CRs, $\gamma$ does not plummet for $k < 1/r'_{L,{\min}}$ (here, $r'_{L,{\min}} = r'_{L,0}$ as $p'_{\min} = p'_0$), as the high-energy CRs are still non-resonant even when the lowest-energy CRs are resonant.
The dispersion relation of NRSI for the isotropic, power-law CR distribution is similar to \citet{bell2004} and \citet{amato2009}.

These results show that the linear growth of NRSI, particularly the dominant modes, is effectively insensitive to the CR distribution.
This highlights a unique characteristic of NRSI, distinguishing it from RSI, where details of the CR distributions play a significant role \citep[see e.g.,][]{holcomb_2019}.

\subsection{Saturation} \label{sec:analytical_saturation}

We now examine the saturation of NRSI.
Being a nonlinear process, deriving an exact formula for the saturated magnetic field from first principles is challenging. 
Recent kinetic studies introduced an anisotropy parameter, $\xi$, to explain and predict the saturated magnetic field for mono-energetic isotropic CRs \citep{gupta_2021,zacharegkas2024modeling}. 
We extend this framework to other CR distributions.

The anisotropy parameter $\xi$ is defined as 
\begin{equation} \label{eq:xi_gupta_2021}
    \xi = \frac{1}{2} \frac{P_{\text{cr,ani}0}}{P_{B0}} \equiv \frac{n_{\text{cr}}p_dv_d}{n_0m_iv^2_{A0}}.
\end{equation}
Here, $P_{\text{cr,ani}0}=n_{\rm cr} p_d v_d$ is the initial anisotropic CR momentum flux\footnote{It is equivalent to the excess anisotropy pressure $P_{\text{cr},xx} - \tfrac{1}{2} \left( P_{\text{cr},yy} + P_{\text{cr},zz} \right)$, where $P_{{\rm cr},ij} \equiv \langle p_{{\rm cr},i} v_{{\rm cr},j} \rangle$ calculated in the plasma ion rest frame. Usually, pressure is evaluated in the comoving frame of the species. However, for NRSI, the free energy is in the CR anisotropy pressure in the plasma-ion rest frame \citep{achterberg2013mirror}, which yields the product $\langle p_{{\rm cr},x}\rangle \langle v_{{\rm cr},x} \rangle = p_dv_d$.}, $P_{B0}=B_0^2/8\pi$ is the initial magnetic energy density; $p_d$ is the mean $x$-momentum, and $v_d$ is the drift speed of CRs, along $B_0 \bm{\hat{x}}$.

We evaluate $v_d$ in the plasma ion rest frame. 
Starting from the CR distributions $g' \left( p,\mu \right)$ in the CR frame, we calculate the mean $x$-component of their velocities in the plasma ion rest frame by applying the boost speed $v_b$ with the CRs. 
This yields
\begin{equation} \label{eq:drift_speed}
    v_d = \int_0^cdv \int_{-1}^1 d\mu \left( \frac{\mu v+v_b}{1+\mu vv_b/c^2} \right) v^2 g'_v(v,\mu),
\end{equation}
where $g'_v(v,\mu) v^2 dv d \mu \equiv g'(p,\mu) p^2 dp d\mu$.
Using Equation \ref{eq:drift_speed}, an approximate expression of $v_d$ for relativistic CRs with $v \approx c$ is 
\begin{equation} \label{eq:drift_vel_approx}
\frac{v_d}{c} \approx \begin{dcases}
        \frac{c}{v_b} - \left( \frac{\frac{c^2}{v_b^2} - 1}{1-\mu'_{\min}} \right) \ln \left| \frac{1+\frac{v_b}{c}}{1+\frac{v_b \mu'_{\min}}{c}} \right| & \left( v_b \neq 0 \right), \\
        \frac{1}{2} \left( 1 + \mu'_{\min} \right) & \left( v_b = 0 \right).
\end{dcases}
\end{equation}
For the isotropic and cone CR distributions, $v_d \approx 2v_b/3$ and $v_d \approx 0.5c$, respectively, for relativistic CRs.

Similar to the drift speed, we calculate the mean momentum $p_d$ from the Lorentz transformation of the CR momenta using the expression
\begin{equation} \label{eq:mean_momentum}
    p_d = \int_0^\infty dp \int_{-1}^1 d\mu \gamma_b \left( \mu p+ Ev_b/c^2 \right) p^2 g'(p,\mu),
\end{equation}
where $\gamma_b=\sqrt{1-\left( v_b/c \right)^2}$ is the Lorentz factor for the boost speed $v_b$ and $E=\sqrt{p^2c^2+m_i^2c^4}$ is the energy.
For relativistic CRs with $p \gg m_ic$, $E \approx pc$; for mono-energetic (ME) and power-law (PL) CRs, Equation \ref{eq:mean_momentum} is approximated as 
\begin{equation} \label{eq:mean_momentum_ME_PL}
\begin{split}
    p_d \approx \gamma_b \left[ \frac{1 + \mu'_{\min}}{2} + \frac{v_b}{c} \right] \times
    \begin{dcases}
        p'_0 & \text{(ME),} \\
        \frac{\ln \left| p'_{\max}/p'_{\min}\right|}{p'^{-1}_{\min} - p'^{-1}_{\max}} & \text{(PL).}
    \end{dcases}
\end{split}
\end{equation}
Using Equation \ref{eq:drift_speed} for $v_d$ and Equation \ref{eq:mean_momentum}  for $p_d$, we calculate $\xi$ from Equation \ref{eq:xi_gupta_2021} for any general CR distribution function boosted with speed $v_b$. 

NRSI saturates when the transverse magnetic field $B_{\perp}$ satisfies the condition 
\begin{equation} \label{eq:xi_vs_saturation}
    \left( \frac{B_{\perp}}{B_0} \right)_{\text{sat}}^2 \approx \frac{1}{4} \frac{P_{\text{cr,ani}0}}{P_{B0}} \implies \left( \frac{B_{\perp}}{B_0} \right)_{\text{sat}} \approx \sqrt{\frac{\xi}{2}}. 
\end{equation}
The factor of $1/4$ arises because the free energy in CR anisotropy pressure is distributed among eight components: three pressure components of plasma ions, three pressure components of plasma electrons (jointly result in plasma heating), and two transverse components of magnetic pressure \citep{zacharegkas2024modeling}.
Equation \ref{eq:xi_vs_saturation} gives the \textit{maximum possible value} of the saturated field for NRSI for any general CR distribution\footnote{We can compare the ``driven" and the ``periodic-box" simulations from \citet{zacharegkas2024modeling} and \citet{gupta_2021}, respectively. 
In both studies, the parameter $\xi$ (or $\xi_0$) is proportional to the initial CR anisotropy pressure, and the saturated (maximum or peak) magnetic field roughly scales with this parameter. 
In both simulation setups, the quenching of the CR current in the background plasma frame causes NRSI to saturate; thus, the same scaling of the saturated magnetic field with the initial anisotropy pressure applies to both.}. 

Our simulations (Section \ref{sec:simulation_saturation}) show that, although Equation \ref{eq:xi_vs_saturation} predicts the saturated field accurately for mono-energetic CRs, it overestimates for power-law CRs. 
For the latter with $p'_{\max} \gg p'_{\min}$, only the low-energy CRs with momenta $p'_{\min} - p'_{\rm eff}$ efficiently contribute to the saturated field, unlike the high-energy CRs with momenta larger than $p'_{\rm eff}$.
Section \ref{sec:simulation_isotropization} shows that $p'_{\rm eff} \approx 9.3 p'_{\min}$ for $f'_{\rm cr}(p) \propto p^{-4}$ (Equation \ref{eq:p_cut}); it decreases for $f'_{\rm cr}(p) \propto p^{-5}$ and increases for $f'_{\rm cr}(p) \propto p^{-3}$.
Using this result, we define an effective anisotropy parameter $\xi_{\rm eff}$, given by
\begin{equation} \label{eq:xi_effective}
\xi_{\rm eff} = 
\begin{dcases}
    \xi[\text{PL:}p'_{\min}-p'_{\rm eff}] & \left( p'_{\max} \geq p'_{\rm eff} \right), \\   \xi[\text{PL:}p'_{\min}-p'_{\max}] & \left( p'_{\max} < p'_{\rm eff} \right),
\end{dcases}
\end{equation}
with momenta ranges $p'_{\min}-p'_{\rm eff}$ and $p'_{\min}-p'_{\max}$, respectively, evaluated using Equation \ref{eq:xi_gupta_2021}. 
Thus, the saturated magnetic field for power-law CRs can be estimated as
\begin{equation} \label{eq:saturation_vs_xi_eff}
\left( \frac{B_{\perp}}{B_0} \right)_{\text{sat,PL}} \approx 
\sqrt{\frac{\xi_{\rm eff}}{2}}.
\end{equation}
Clearly, $\xi_{\rm eff}=\xi$ for mono-energetic CRs, and Equation \ref{eq:saturation_vs_xi_eff} reduces to Equation \ref{eq:xi_vs_saturation}.
The predictions hold for both relativistic and nonrelativistic CRs.

\section{Numerical setup} \label{sec:numerical_setup}

We use a publicly available version of the electromagnetic particle-in-cell code, Tristan-MP \citep{spitkovsky2005simulations}, modified to investigate the CR streaming instability \citep{gupta_2021,lichko_2025}.
Our simulations are spatially in 1D, with velocities and fields having all three components.

We initialize the grids with a magnetic field $\bm{B}=B_0 \bm{\hat{x}}$, the background plasma species (ions and electrons) with thermal distributions, and the CR ions with mono-energetic or power-law momentum distributions.
For isotropic distributions, we boost each CR particle with a velocity $v_b \bm{\hat{x}}$, unlike the cone distributions, where we do not boost the CRs.
We boost the thermal electrons to balance the CR current (see Equation \ref{eq:current_nutral}).
Starting from this initial setup, we solve Maxwell's equations to evolve the electromagnetic fields and the equation of motion to evolve the particles' positions at each iteration, under periodic boundary conditions.
Thus, the system evolves self-consistently.
Table \ref{tab:grid_field_particle} contains the parameters of our numerical setup.

\begin{table}
  \begin{threeparttable}
    \caption{Details of grid, initial field, and particles}
    \centering
     \begin{tabularx}{\linewidth}{lr}
        \toprule
        \hline
        Parameters & Values \\
        \midrule
        \hline
        Box size $\left( N_x \times N_y \right)$ & $300,000 \times 2$ \\
        \{for benchmark runs\} & $1,000,000 \times 2$ \\
        Cell width ($\Delta x$) & $0.2 d_e$ \\
        Time step of iteration ($\Delta t$) & $0.09 \omega_{pe}^{-1}$ \\
        \midrule
        Alfv\'en speed $(v_{A0}=B_0/\sqrt{4 \pi m_in_0})$ & $0.02c$ \\
        \midrule 
        Particle-per-cell (each species) & $25$ \\
        Mass ratio $(m_i/m_e)$ & 25 \\
        Ion thermal speed $(v_{\text{th},i}=\sqrt{k_{\rm B} T_i/m_i})$ & $0.008c$ \\
        Temperature ratio $(T_i/T_e)$ & $1$ \\
        \bottomrule
     \end{tabularx} \label{tab:grid_field_particle}
    \begin{tablenotes}
      \small
      \item \textbf{Note:} Plasma frequency and skin depth of background electrons are $\omega_{pe}$ and $d_e$. 
      Changing the box size, $\Delta x$, $\Delta t$, particle-per-cell, and mass ratio yields consistent results.
    \end{tablenotes}
  \end{threeparttable}
\end{table}

We use an artificial ion-to-electron mass ratio $m_i/m_e$ to reduce their scale separation, which has almost no effect in the saturation of the magnetic field \citep[see, e.g., Appendix A in ][]{gupta+24b}.
We implement the effective number densities of the species by adjusting their masses, keeping their charge-to-mass ratios intact. 
By fixing $n_e=1$, we fix a $n_{\text{cr}}/n_e$ for a particular simulation, and set $n_i/n_e$ as per Equation \ref{eq:charge_nutral}. 
We perform simulations, detailed in Table \ref{tab:sims_data}, to primarily answer the questions listed in Section \ref{sec:intro}.

\begin{table}
  \begin{threeparttable}
    \caption{Details of parameters for different runs}
    \centering
     \begin{tabularx}{\linewidth}{llccccc}
        \toprule
        \hline
        Sl. & Runs & $v_d$ & $p_d$ & $\xi$ & $\xi_{\rm eff}$ & $k_{\rm fast}d_i$ \\
        No. & & $[c]$ & $[m_ic]$ & & & $(\times 10^{-2})$ \\ 
        \midrule
        \hline
        I1 & MEI:5$\star$ & 0.3 & 2.2 & 15 & 15 & 7.03 \\
        I2 & PLI:5-30     & 0.3 & 4.7 & 32 & 32 & 6.95 \\
        I3 & PLI:5-80     & 0.3 & 6.4 & 44 & 38 & 6.95 \\
        I4 & PLI:5-400$\star$ & 0.3 & 9.6 & 66 & 38 & 6.94 \\
        I5 & PLI:10-400   & 0.3 & 16.5 & 114 & 75 & 6.91 \\
        I6 & PLI:20-400  & 0.3 & 27.5 & 190 & 150 & 6.90 \\
        I7 & MEI:10.7     & 0.3 & 4.7 & 32 & 32 & 6.93 \\
        I8 & MEI:14.7     & 0.3 & 6.4 & 44 & 44 & 6.91 \\
        I9 & MEI:22       & 0.3 & 9.6 & 66 & 66 & 6.90 \\
        \midrule
        C1 & MEC:5$\star$ & 0.5 & 2.5  & 30 & 30 & 12.25 \\
        C2 & PLC:5-30     & 0.5 & 5.3  & 66 & 66 & 12.40 \\
        C3 & PLC:5-80     & 0.5 & 7.3  & 90 & 76 & 12.41 \\
        C4 & PLC:5-400$\star$ & 0.5 & 10.9 & 136 & 76 & 12.41 \\
        C5 & PLC:10-400   & 0.5 & 18.9 & 235 & 155 & 12.48 \\
        C6 & PLC:20-400  & 0.5 & 31.5 & 394 & 312 & 12.49 \\
        C7 & MEC:10.7     & 0.5 & 5.3  & 66 & 66 & 12.45 \\
        C8 & MEC:14.7     & 0.5 & 7.3  & 91 & 91 & 12.47 \\
        C9 & MEC:22       & 0.5 & 11.0 & 137 & 137 & 12.49 \\
        \midrule
        I10 & PLI:50-400:$10^{-3}$    & 0.3 & 51.9 & 36 & 36 & 0.69 \\
        C10 & PLC:50-400:$10^{-3}$    & 0.5 & 59.4 & 74 & 74 & 1.25 \\
        \bottomrule
     \end{tabularx} \label{tab:sims_data}
    \begin{tablenotes}
      \small
      \item \textbf{Note:}
      In the labels of the runs, `ME' and `PL' refer to mono-energetic and power-law CR momentum distributions, and `I' and `C' denote isotropic or cone pitch-angle distributions, respectively.  
      The numbers following the colon (:) are $p'_0$ for mono-energetic CRs, or $p'_{\text{min}}-p'_{\text{max}}$ for power-law CRs.
      The CR number fraction $n_{\text{cr}}/n_0=10^{-2}$ for all these runs, except I10 and C10, for which $n_{\text{cr}}/n_0=10^{-3}$ (mentioned in the label).
      The boost speeds for isotropic and cone CRs are $\bm{v_b}=0.4c \bm{\hat{x}}$ and $\bm{v_b}=\bm{0}$, respectively.
      Remaining columns show drift speed $v_d$, mean momentum $p_d$, anisotropy parameter $\xi$, effective anisotropy parameter $\xi_{\rm eff}$, and fastest growing mode $k_{\text{fast}}d_i \left( = \gamma_{\text{fast}}/\omega_{ci} \right)$. 
      Runs marked with ($\star$) are the benchmark runs; movies for them are provided (\href{https://www.youtube.com/playlist?list=PLsIECy7LXbWI22xY_-57BXBjiLnbcP1mM}{Click here}).
    \end{tablenotes}
  \end{threeparttable}
\end{table}

\section{Results} \label{sec:results}

We study the linear regime (Section \ref{sec:simulation_linear_growth}) and the nonlinear regime (Section \ref{sec:simulation_saturation}), particularly the saturation, of NRSI.
We focus on how these regimes differ between mono-energetic and power-law CR distributions.

\subsection{Linear Growth: Mono-energetic vs. Power-law} \label{sec:simulation_linear_growth}

We investigate the spatial profiles of the magnetic fields (Section \ref{sec:simulation_spatial_profile}) and their Fourier components (Section \ref{sec:simulation_Fourier_analysis}).
We show that, in the linear phase, the instability is largely independent of the CR distributions.

\subsubsection{Spatial Profiles of Magnetic Field Components} \label{sec:simulation_spatial_profile}

\begin{figure*}
    \centering
    \begin{minipage}{0.61\textwidth}
        \centering
        \includegraphics[width=\linewidth]{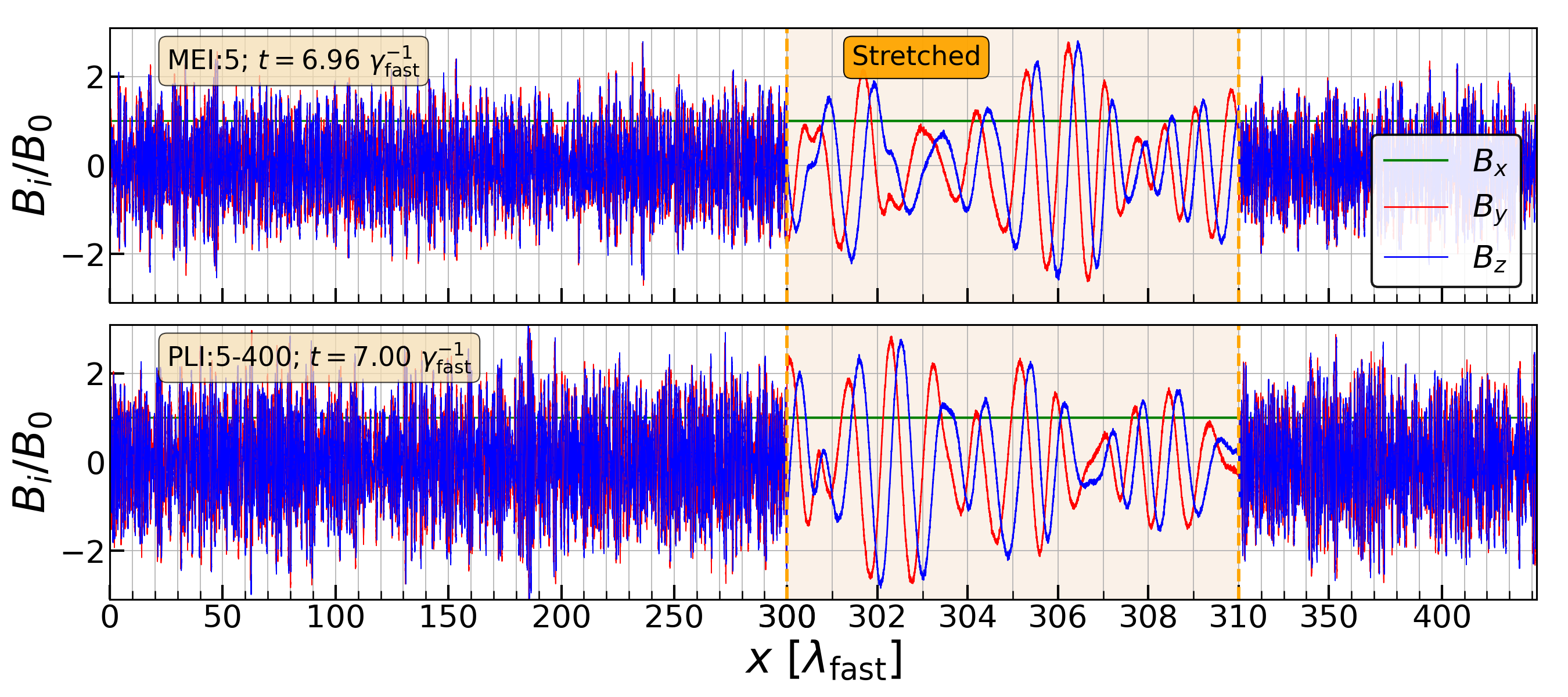}
    \end{minipage}%
    \hspace{0.02\textwidth}%
    \begin{minipage}{0.37\textwidth}
        \centering
        \includegraphics[width=\linewidth]{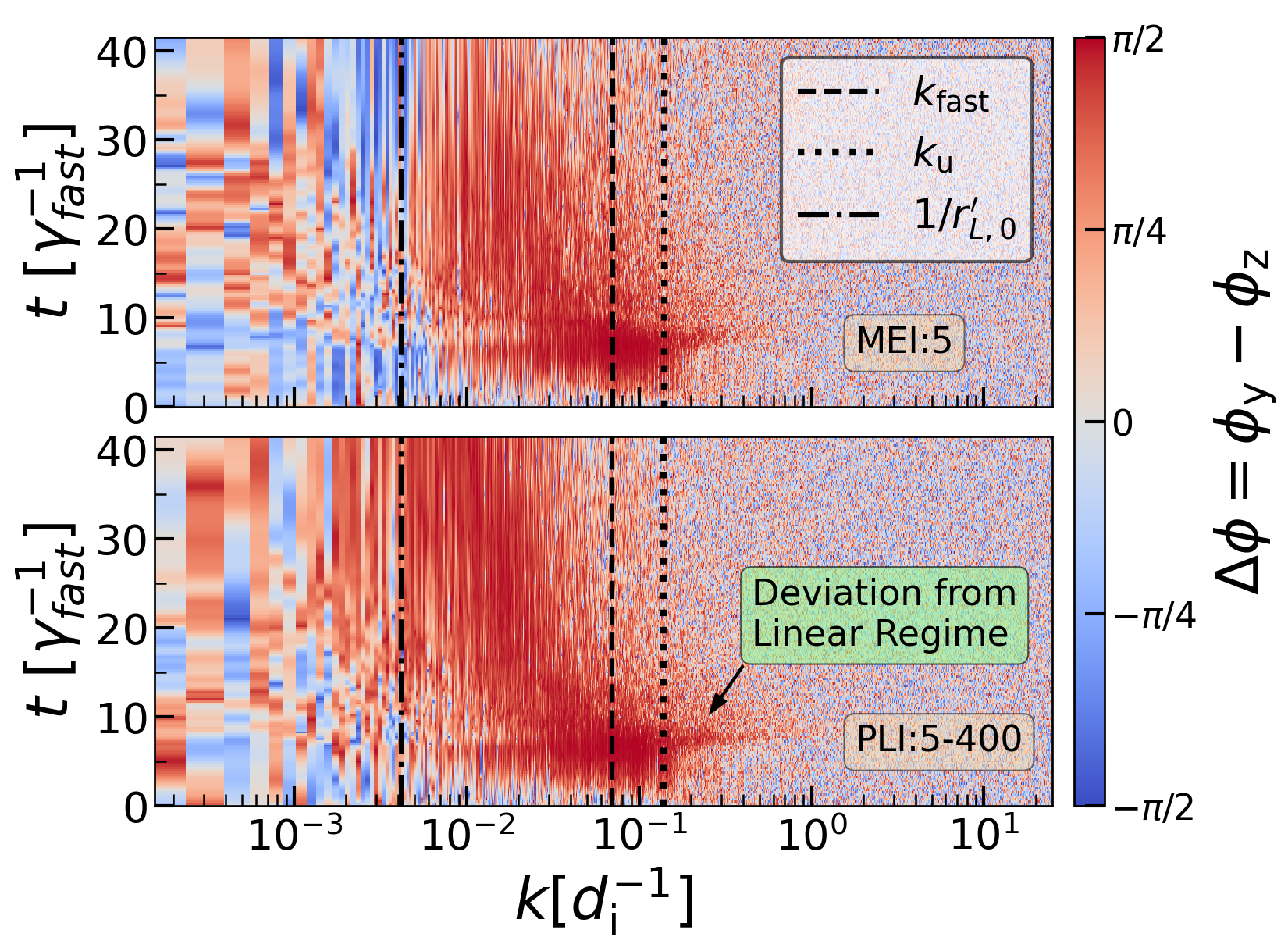}
    \end{minipage}
    \caption{Diagnostics of the magnetic fields for a mono-energetic (upper panels) and a power-law (lower panels) isotropic CR distributions.
    [Left panels] The spatial profiles of the magnetic field components, normalized to the initial magnetic field $B_0$, at $t \approx 7\gamma_{\text{fast}}^{-1}$. 
    The $x$-axes are normalized to the wavelengths of the fastest-growing modes $(\lambda_{\text{fast}}=2 \pi /k_{\text{fast}} \approx 90 d_i)$. 
    From the shaded zoomed-in part, the dominant wavelengths, almost equal to $\lambda_{\text{fast}}$, are prominent. 
    [Right panels] Time evolution of helicities $\Delta \phi (k)$. 
    The positive and negative values signify R-handed NRSI and L-handed RSI, respectively.} 
    \label{fig:MagFld_Snaps_pcr_5}
\end{figure*}


\begin{figure*}
    \centering
    \begin{minipage}{0.61\textwidth}
        \centering
        \includegraphics[width=\linewidth]{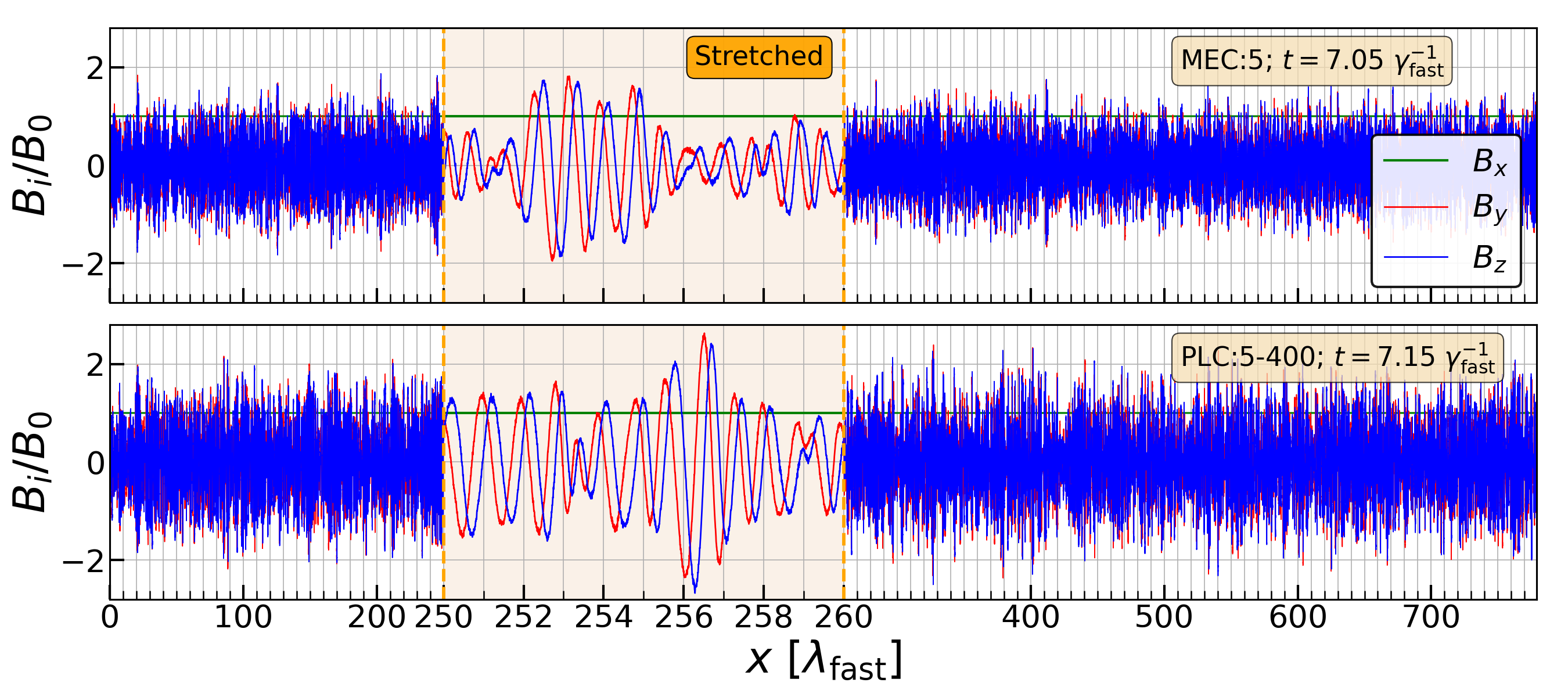}
    \end{minipage}%
    \hspace{0.02\textwidth}%
    \begin{minipage}{0.37\textwidth}
        \centering
        \includegraphics[width=\linewidth]{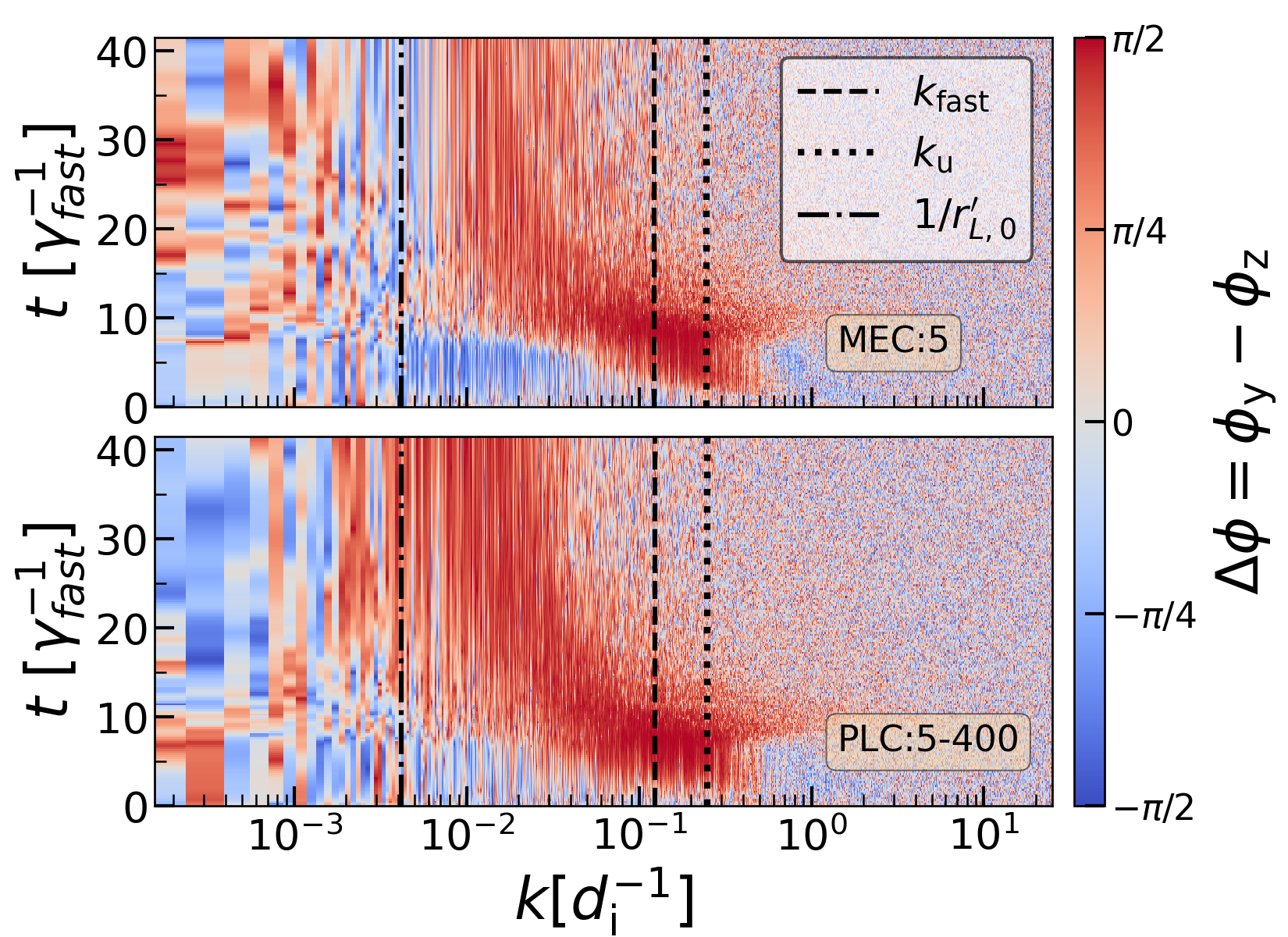}
    \end{minipage}
    \caption{Similar to Figure \ref{fig:MagFld_Snaps_pcr_5}, except that the CRs initially have cone distributions.
    [Left panels] The spatial profiles of the magnetic field components at $t \approx 7\gamma_{\text{fast}}^{-1}$. 
    The $x$-axes are normalized to $\lambda_{\text{fast}}=2 \pi /k_{\text{fast}} \approx 50 d_i$. 
    In the shaded zoomed-in part, the dominant wavelengths $\sim \lambda_{\text{fast}}$ are prominent. 
    [Right panels] Time evolution of helicities $\Delta \phi (k)$.}
    \label{fig:MagFld_snaps_HC}
\end{figure*}


We plot the magnetic field components $B_x$, $B_y$, and $B_z$ for the benchmark isotropic runs MEI:5 and PLI:5-400 (I1 and I4 in Table \ref{tab:sims_data}).
The left panels of Figure \ref{fig:MagFld_Snaps_pcr_5} show their profiles at $t \approx 7 \gamma_{\text{fast}}^{-1}$, a time during the linear growth.
Since the entire box is large (to capture the Larmor radius | hence, the resonant mode | of the highest-energy CR, $r'_{L, p'=400m_ic}\approx 1.95 \times 10^4 d_i$), we zoom into $x \in [300,310] \lambda_{\text{fast}}$. 
The profiles of $B_y$ and $B_z$ at $t \approx 7 \gamma_{\text{fast}}^{-1}$ show their growth, as $B_y/B_0$ and $B_z/B_0 \gtrsim 1$. 
The dominant wavelengths $\sim \lambda_{\text{fast}}$ are due to NRSI.
The profiles are similar for both runs.

When driven by CR ions, NRSI causes the transverse components to orient anticlockwise along the background field.
Thus, $B_y(x)$ lags behind $B_z(x)$ by a phase difference $\Delta \phi \sim \pi/2$, and we identify the components as {\it right-handed} (R-handed).
In contrast, {\it left-handed} (L-handed) RSI causes the transverse components to orient clockwise, and $B_y(x)$ leads $B_z(x)$ with $\Delta \phi \sim - \pi/2$ \citep{gupta_2021}.
In the left panels of Figure \ref{fig:MagFld_Snaps_pcr_5}, $B_y$ lags behind $B_z$ by $\Delta \phi \sim \pi/2$, due to R-handed NRSI.

\subsubsection{Fourier Mode Analysis} \label{sec:simulation_Fourier_analysis}

We perform Fourier analysis of the transverse field components to identify different NRSI modes.
$\tilde{B}_y (k)$ and $\tilde{B}_z (k)$ are the Fourier transforms of $B_y (x)$ and $B_z (x)$, and we use them to calculate {\it helicity} of the modes using Equations \ref{eq:helicity} and \ref{eq:stokes_from_fourier} (see Appendix \ref{sec:Helicity} for details).
Helicity is {\it equivalent} to the phase difference in the $k$-space, $\Delta \phi (k) = \phi_y (k) - \phi_z (k)$. 
The positive and negative values of helicity, with $\Delta \phi (k) \sim \pi/2$ and $-\pi/2$, signify R-handed NRSI and L-handed RSI, respectively.

The right panels of Figure \ref{fig:MagFld_Snaps_pcr_5} show the helicities for the benchmark runs.
We indicate $k_{\text{fast}}$, the upper cutoff of NRSI $k_{\text{u}}=2k_{\text{fast}}$, and the mode resonant with CRs $k = 1/r'_{L,0} = 1/r'_{L,\min}$. 
For both runs, at $t \lesssim 10 \gamma^{-1}_{\text{fast}}$, the R-handed NRSI with $k \sim k_{\text{fast}}$ is prominent; modes with $1/r'_{L,0} \lesssim k \lesssim k_{\text{u}}$ show positive helicity.
Later at $t \gtrsim 10 \gamma^{-1}_{\text{fast}}$, for MEI:5, $k \lesssim 1/r'_{L,0}$ (e.g., $k \in (1,4)\times10^{-3}d_i^{-1}$) show negative helicity due to RSI. 
In contrast, for PLI:5-400, these modes show positive helicity due to NRSI, consistent with the analytical dispersion relation (see Figure \ref{fig:dispersion_analytical}).

A few modes with $k>k_{\text{u}}$ showing positive helicity at $t \gtrsim 10 \gamma_{\text{fast}}^{-1}$, annotated in the plot, is a nonlinear phenomenon. 
As $B_{\perp}/B_0$ becomes larger than 1, small-scale structures emerge due to turbulence, which cause these larger $k$-modes to show positive helicity.
Due to $B_{\perp}/B_0 \gtrsim 1$, linear theory is no longer applicable, and the slight deviation is therefore expected.

In Figure \ref{fig:MagFld_snaps_HC}, we perform the diagnostics for the benchmark cone runs, MEC:5 and PLC:5-400 (C1 and C4 in Table \ref{tab:sims_data}), and obtain similar results.
We also roughly calculate the growth rates of the R-handed modes, and compare them with the analytical expectations, for these benchmark runs in Appendix \ref{sec:NRSI_Growth_Rates_Sims}.

\subsubsection{Evolution of Transverse Magnetic Fields} \label{sec:simulation_evolution_of_BPerp}

\begin{figure}
    \centering
    \includegraphics[width=0.47\textwidth]{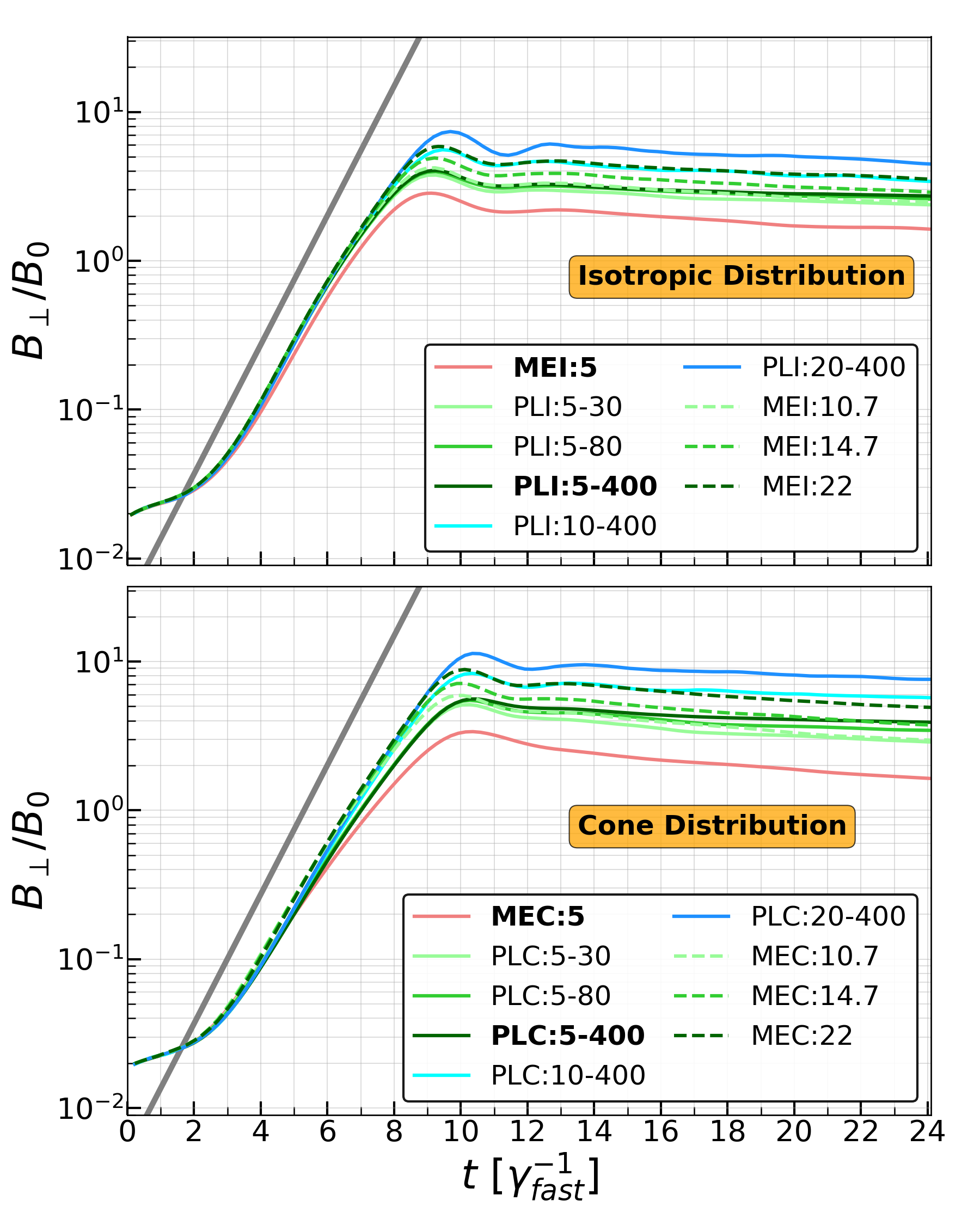}
    \caption{Time evolution of the transverse magnetic fields $B_{\perp}$ for the runs in Table \ref{tab:sims_data}.
    Their linear growth shows good agreement with the analytical prediction (grey solid lines).
    However, the saturated magnetic fields are different.}
    \label{fig:MagFld_growth}
\end{figure}

Figure \ref{fig:MagFld_growth} shows the evolution of the transverse magnetic fields $B_\perp = \sqrt{B_y^2 + B_z^2}$ for the runs with isotropic (I1-I9) and cone (C1-C9) CRs in Table \ref{tab:sims_data}. 
In the linear regime ($B_\perp /B_0 \lesssim 1$), $B_\perp$ grows exponentially for all runs, consistent with the analytical predictions. 
In the nonlinear regime ($B_{\perp}/B_0 \gtrsim 1$) at time $t \sim 10 \gamma^{-1}_{\text{fast}}$, $B_{\perp}$ stops growing and the instability saturates. 
The saturated magnetic fields differ for the runs.

\subsection{Saturation: Mono-energetic vs. Power-law} \label{sec:simulation_saturation}

We test the existing saturation prescription and model the saturated magnetic fields (Section \ref{sec:modelling_simulation_saturation}).
We investigate why saturation levels differ across CR distributions with similar $\xi$ (Section \ref{sec:saturation_vs_CRdistribution}), and examine CR isotropization (Section \ref{sec:simulation_isotropization}).

\subsubsection{Modeling the Saturated Magnetic Field}  \label{sec:modelling_simulation_saturation}

\begin{figure*}
    \centering
    \includegraphics[width=0.99\textwidth]{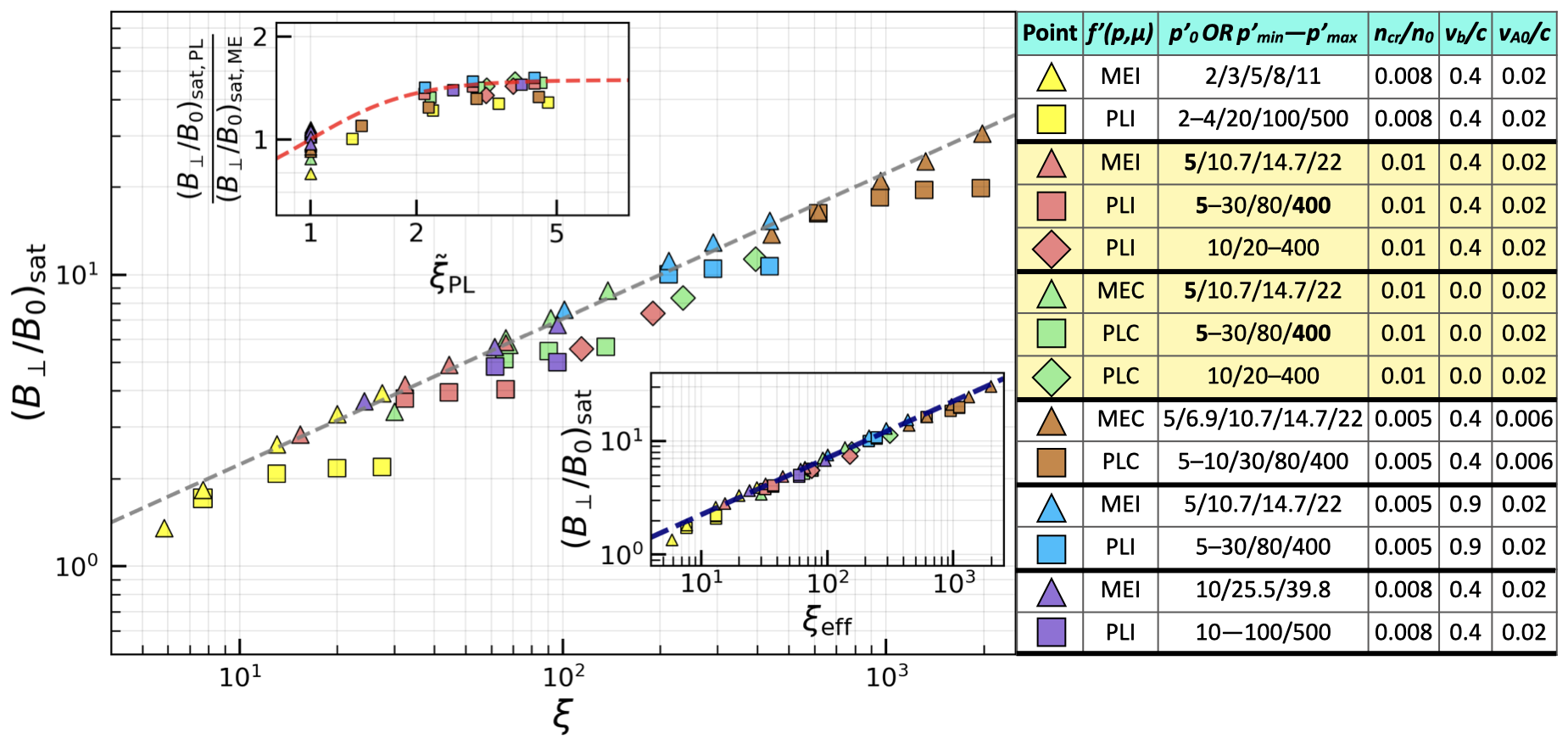}
    \caption{The saturated magnetic field $\left( B_{\perp}/B_0 \right)_{\rm sat}$ versus the anisotropy parameter $\xi$.
    The shapes and colors represent different simulations (see the table on the right; see Table \ref{tab:sims_data} for the nomenclature of the runs).
    Momenta separated by `/' represent different runs (e.g. PLI:5-30/80 implies PLI:5-30 and PLI:5-80).
    For the points with the same colors and shapes, larger $\xi$'s correspond to larger momenta (e.g. for PLI:5-30/80, $\xi$ for PLI:5-80 is larger than PLI:5-30).
    The grey dashed line represents the prediction, i.e. Equation \ref{eq:xi_vs_saturation}.
    The runs with mono-energetic CRs mostly follow this prediction, unlike those with power-law CRs.
    Upper Inset: Modelling of $\left( B_{\perp}/B_0 \right)_{\rm sat}$ using Equation \ref{eq:xi_vs_saturation_fitting} (red dashed line).
    Lower Inset: Modelling of $\left( B_{\perp}/B_0 \right)_{\rm sat}$ using Equation \ref{eq:saturation_vs_xi_eff} (navy-blue dashed line). }
    \label{fig:saturation_vs_xi}
\end{figure*}

We take the maximum values of $B_{\perp}/B_0$ (e.g. in Figure \ref{fig:MagFld_growth}) as the saturated magnetic field $\left( B_{\perp}/B_0 \right)_{\text{sat}}$.
Figure \ref{fig:saturation_vs_xi} shows $\left( B_{\perp}/B_0 \right)_{\text{sat}}$ versus $\xi$ for our runs in Table \ref{tab:sims_data}, in addition to other runs.
We also include boosted cone CR distributions (brown points).

First, we investigate the CR distributions with different $\xi$'s.
We consider the pink symbols: triangles for mono-energetic CRs, and squares and diamonds for power-law CRs. 
For mono-energetic CRs, as $p'_0$ increases, $\xi$ increases (see Equations \ref{eq:mean_momentum_ME_PL} and \ref{eq:xi_gupta_2021}), and $\left( B_{\perp}/B_0 \right)_{\text{sat}}$ increases accordingly. 
For power-law CRs, MEI:5 and PLI:5-30/80/400\footnote{MEI:5 is a special instance of power-law CRs with $p'_{\min}=p'_{\max}=5m_ic$; $p'_{\max}$ increases gradually for these runs.}, as $p'_{\max}$ increases, $\xi$ increases (see Equation \ref{eq:mean_momentum_ME_PL}). 
However, $\left( B_{\perp}/B_0 \right)_{\text{sat}}$ increases from MEI:5 to PLI:5-30, and it stops increasing significantly in PLI:5-80 and PLI:5-400. 
In contrast, for the runs PLI:5/10/20-400 with increase in $p'_{\min}$, $\left( B_{\perp}/B_0 \right)_{\text{sat}}$ increases prominently.
Thus, for power-law CRs with a wide energy range and a fixed $n_{\rm cr}$, $\left( B_{\perp}/B_0 \right)_{\text{sat}}$ is sensitive to a change in $p'_{\min}$, but does not depend strongly on a change in $p'_{\max}$.

Next, we investigate the CR distributions with similar $\xi$'s.
For them, $\left( B_{\perp}/B_0 \right)_{\text{sat}}$'s are not similar, unlike what is predicted in Equation \ref{eq:xi_vs_saturation}.
$\left( B_{\perp}/B_0 \right)_{\text{sat}}$'s are similar when the difference between $p'_{\text{min}}$ and $p'_{\text{max}}$ for the power-law CRs is smaller, e.g., MEI:10.7 and PLI:5-30 with $\xi \sim 32$.
When this difference is larger, $\left( B_{\perp}/B_0 \right)_{\text{sat}}$ for the power-law CRs is smaller than the mono-energetic CRs, e.g., MEI:22 and PLI:5-400 with $\xi \sim 66$.
We observe similar results across all the runs, including the cone CRs (green points) and the boosted-cone CRs (brown points). 

The mono-energetic CRs show good agreement with Equation \ref{eq:xi_vs_saturation}, unlike the power-law CRs, for which $\xi$ overestimates $\left( B_{\perp}/B_0 \right)_{\text{sat}}$.
We implement a fitting function to model their saturated fields
by introducing an effective cutoff momentum $p'_{\rm eff}$ (Equation \ref{eq:p_cut}), which is used to evaluate the effective anisotropy pressure $\xi_{\rm eff}$ (Equation \ref{eq:xi_effective}) and the saturated field for power-law CRs (Equation \ref{eq:saturation_vs_xi_eff}).

For the power-law CRs with a fixed $p'_{\text{min}}$ and an increasing $p'_{\text{max}}$ (e.g. MEI:5 and PLI:5-30/80/400), $\left( B_{\perp} / B_0 \right)_{\text{sat}}$ vs $\xi$ show a `$\tanh$' dependence in $\log$-$\log$ space.
This motivates a fitting function for power-law (PL) CRs with anisotropy parameter $\xi[\text{PL:}p'_{\text{min}}-p'_{\text{max}}]$, given by
\begin{equation} \label{eq:xi_vs_saturation_fitting}
\left( \frac{B_{\perp}}{B_0} \right)_{\text{sat,PL}} = \left( \frac{B_{\perp}}{B_0} \right)_{\text{sat,ME}} \exp \left(0.4 \frac{ \tilde{\xi}_{\text{PL}} ^3 - 1}{ \tilde{\xi}_{\text{PL}}^3 + 1} \right).
\end{equation}
Here, $(B_{\perp}/B_0)_{\text{sat,ME}}$ and $\tilde{\xi}_{\text{PL}}$ are defined as
\begin{subequations} \label{eq:xi_tilda}
    \begin{align}
        &\left( \frac{B_{\perp}}{B_0} \right)_{\text{sat,ME}} = \sqrt{\frac{\xi[\text{ME:}p'_{\text{min}}]}{2}} \text{ and}\\
        &\tilde{\xi}_{\text{PL}} = \frac{\xi[\text{PL:}p'_{\text{min}}-p'_{\text{max}}]}{\xi[\text{ME:}p'_{\text{min}}]},
    \end{align}
\end{subequations}
respectively.
$\xi[\text{ME:}p'_{\text{min}}]$ is anisotropy parameter $\xi$ calculated for mono-energetic (ME) CRs with momentum $p'_{\min}$ and number density $n_{\rm cr}$.
The fitting function indicates that the saturated magnetic field depends on both the anisotropy parameter and the minimum momentum of the power-law CRs.
Also, for $p'_{\text{max}} \gg p'_{\text{min}}$, $\tilde{\xi}_{\text{PL}}^3 \gg 1$; hence, Equation \ref{eq:xi_vs_saturation_fitting} yields $(B_{\perp}/B_0)_{\text{sat,PL}} \approx \exp(0.4) \times \sqrt{\xi[\text{ME:}p'_{\text{min}}]/2}$.
This is the maximum saturated magnetic field for power-law relativistic CRs (with $f'_{\text{cr}}(p) \propto p^{-4}$) with a wide energy range.
Equation \ref{eq:xi_vs_saturation_fitting} also applies to mono-energetic CRs, for which $p'_{\text{min}}=p'_{\text{max}}=p'_0$ and $\tilde{\xi}_{\text{PL}} = 1$; hence, Equation \ref{eq:xi_vs_saturation_fitting} reduces to Equation \ref{eq:xi_vs_saturation}.

The upper inset in Figure \ref{fig:saturation_vs_xi} shows the normalized saturated magnetic field $\left( B_{\perp}/B_0 \right)_{\text{sat,PL}} / \left( B_{\perp}/B_0 \right)_{\text{sat,ME}}$ versus the normalized anisotropy parameter $\tilde{\xi}_{\text{PL}}$.
The runs align well with the fit, i.e. Equation \ref{eq:xi_vs_saturation_fitting}.

\subsubsection{Mono-energetic vs Power-law CRs with similar $\xi$} \label{sec:saturation_vs_CRdistribution}

\begin{figure} 
    \centering
    \includegraphics[width=0.47\textwidth]{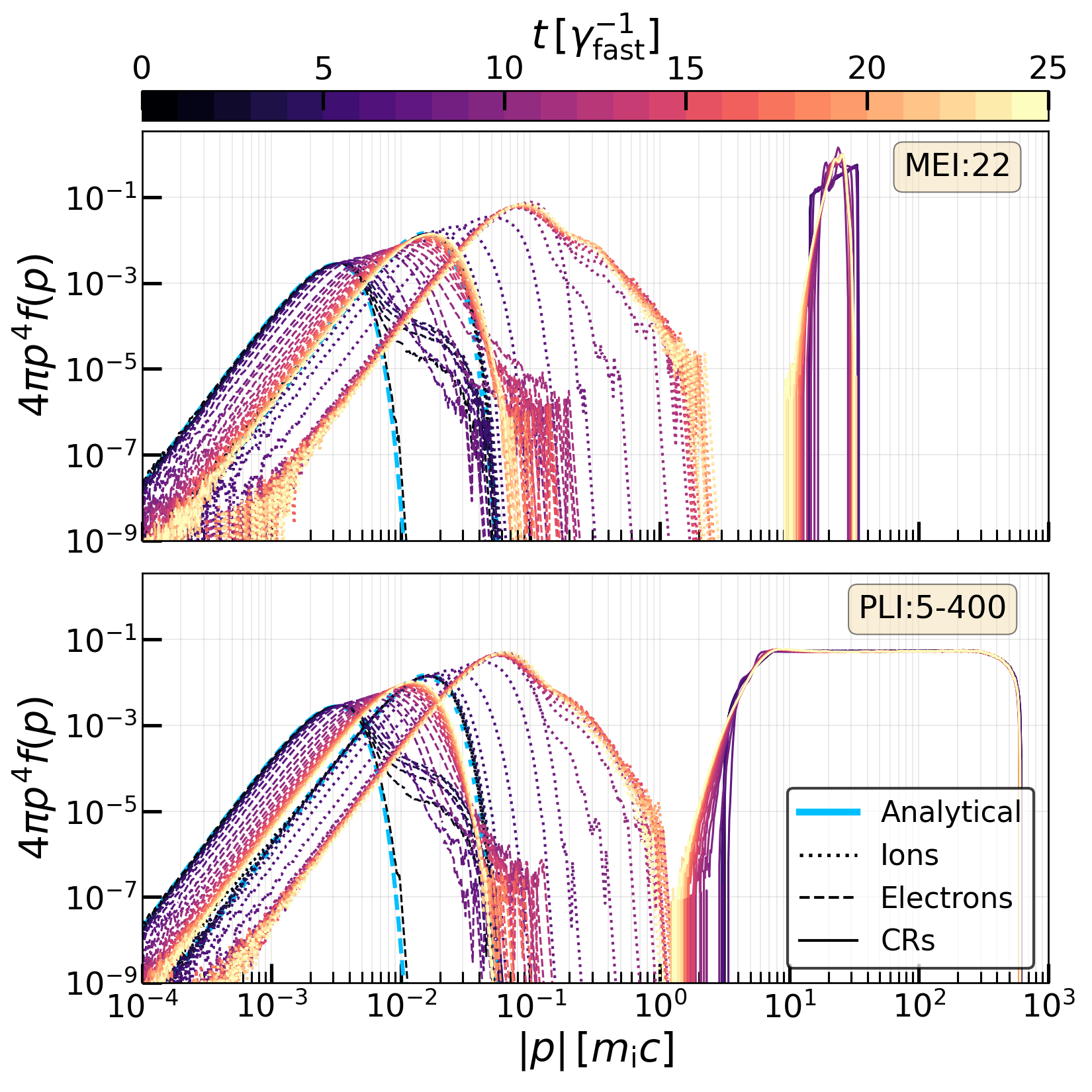}
    \caption{The momentum distributions of plasma ions (dotted lines), plasma electrons (dashed lines), and CRs (solid lines) in the plasma ion rest frame, for mono-energetic (top) and power-law (bottom) CRs with similar $\xi \approx 66$.
    The colors represent distributions at different times.
    Heating of plasma species (nonthermal tails and shifts in thermal peaks) and changes in CR distributions are observed at later times.
    The higher-energy CRs in the power-law distribution are essentially undisturbed. }
    \label{fig:momdis_comparison_HI}
\end{figure}

For the similar $\xi$ (or similar initial CR anisotropy pressure; see Equation \ref{eq:xi_gupta_2021}), the saturated magnetic field for the mono-energetic CRs is larger than the power-law CRs (see Section \ref{sec:modelling_simulation_saturation}).
The saturation of NRSI is caused by the isotropization of CRs, through which the free energy in the CR anisotropy pressure gets transferred to the transverse magnetic fields and the background plasma.
We show that the ineffective isotropization of higher-energy CRs for a power-law distribution leads to a smaller saturated field compared to mono-energetic CRs.
We consider two runs, MEI:22 and PLI:5-400, with $\xi \approx 66$ (I9 and I4 in Table \ref{tab:sims_data}), and analyze the differences in the evolution of their momentum distributions and pressure components.

Figure \ref{fig:momdis_comparison_HI} shows the momentum distributions for the two runs in the plasma-ion rest frame. 
For the plasma ions and electrons, we observe the emergence of nonthermal tails and the shifts in their thermal peaks, due to plasma heating.
For the CRs, the entire distribution for MEI:22 changes over time.
In contrast, for PLI:5-400, CRs with the lowest momenta are affected, without any significant change in the high-energy CRs.

\begin{figure} 
    
    \centering
    \includegraphics[width=0.47\textwidth]{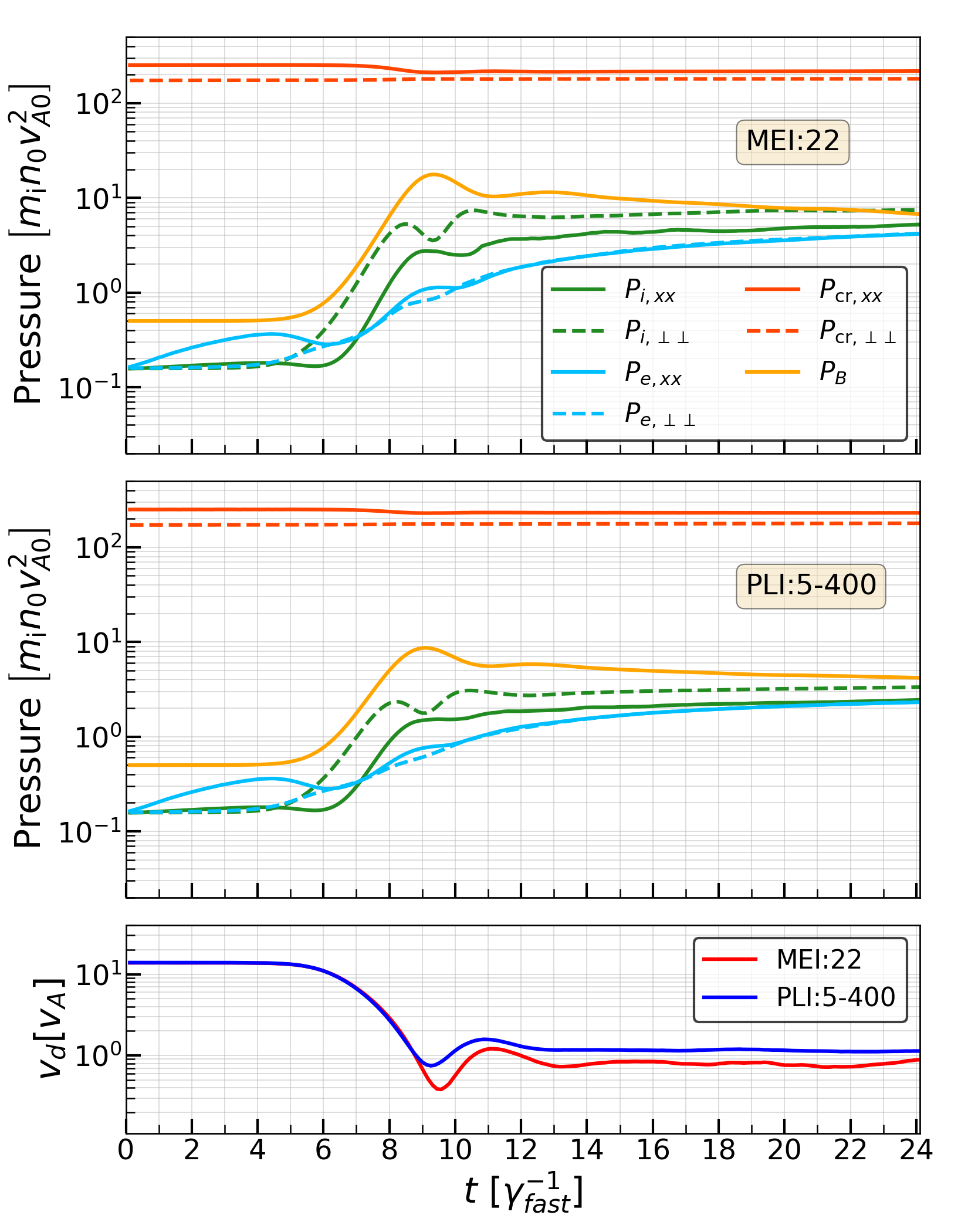}
    \caption{[Top two panels] Evolution of the longitudinal $\left( P_{xx} \text{, solid lines} \right)$ and transverse $\left( P_{\perp \perp} \text{, dashed lines} \right)$ pressure of plasma ions (green), plasma electrons (cyan), and CRs (red) in plasma ion rest frame, for the mono-energetic (top) and power-law (middle) CRs with a similar $\xi \approx 66$.
    The net magnetic pressure (orange) is $P_B=B^2/8\pi$, where $B^2=B_0^2+B_{\perp}^2$.
    After saturation, isotropization of CR pressure components for MEI:22 is apparent, whereas they remain anisotropic for PLI:5-400.
    [Bottom panel] CR drift $v_d$ in plasma ion rest frame, normalized to the Alfv\'en speed $v_A$.
    After saturation, sub-Alfv\'enic CR drifts cease further magnetic field growth.}
    \label{fig:pressure_comparison_HI}
\end{figure}

Figure \ref{fig:pressure_comparison_HI} shows longitudinal $\left( P_{xx} \right)$ and transverse $\left( P_{\perp \perp} = \left( P_{yy} + P_{zz} \right) /2 \right)$ pressure components for plasma species and CRs, along with the magnetic pressure, for the two runs.
We compute the pressure component $P_{s,ij}=\langle p_{s,i} v_{s,j} \rangle$ of species $s$ in the plasma-ion rest frame.
The pressures of plasma ions and electrons evolve similarly in both runs. 
However, the heating of plasma species, as well as the saturated magnetic pressure, is larger for MEI:22 than PLI:5-400 at $t \sim 9 \gamma_{\text{fast}}^{-1}$, when the overall saturation is reached.
The increase in magnetic and background plasma pressures results from the equipartition of the CR anisotropy pressure $P_{\text{cr},xx} - P_{\text{cr},\perp \perp}$.
For MEI:22, CRs isotropize efficiently; thus, $P_{\text{cr},xx} \approx P_{\text{cr},\perp \perp}$ at $t \sim 9 \gamma_{\text{fast}}^{-1}$, maximally amplifying the magnetic and background plasma pressures.
In contrast, for PLI:5-400 at $t \sim 9 \gamma_{\text{fast}}^{-1}$, the CR pressure components remain anisotropic, particularly due to the slower isotropization of the high-energy CRs (see Section \ref{sec:simulation_isotropization}); it leads to a smaller magnetic and background plasma pressures. 
Later at $t \gtrsim 9 \gamma_{\rm fast}^{-1}$, the CR pressure components of the high-energy particles continue to isotropize slowly, but the magnetic field gradually decays with time. 
The residual anisotropy pressure of high-energy CRs contributes to the gradual heating of the background plasma instead of feeding the transverse fluctuations, due to the continuous damping of the magnetic energy.

The transverse field growth is stopped, as the net CR drift does not remain super-Alfv\'enic after the overall saturation. 
In the bottom panel of Figure \ref{fig:pressure_comparison_HI}, $v_d/v_A \sim 1$ at $t \gtrsim 9 \gamma_{\rm fast}^{-1}$ for both runs.
Even if the high-energy CRs maintain a super-Alfv\'enic drift, the isotropization of more numerous lower-energy CRs causes the overall drift relaxation.

\subsubsection{Isotropization of CRs} \label{sec:simulation_isotropization}

\begin{figure} 
    
    \centering
    \includegraphics[width=0.47\textwidth]{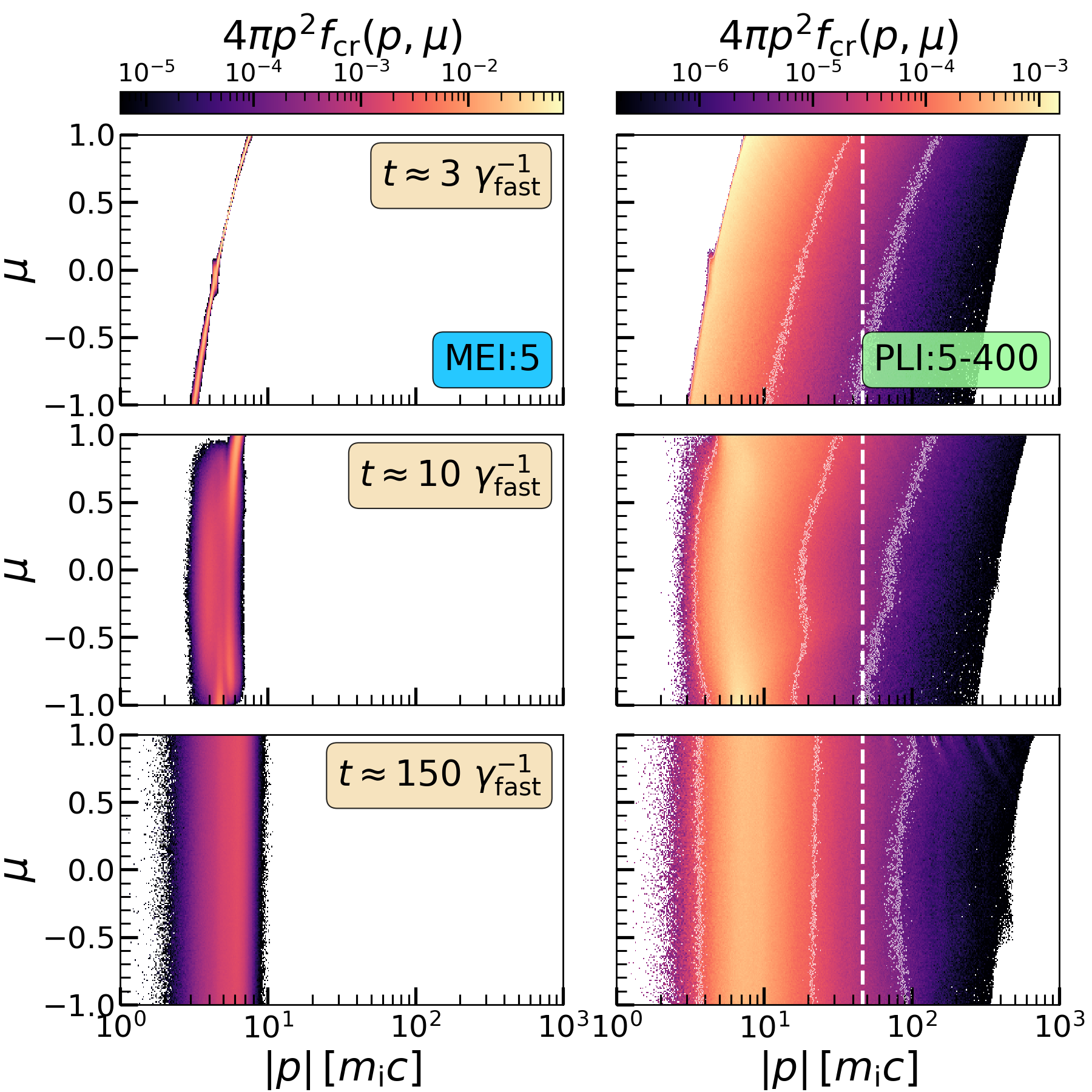}
    \caption{CR distributions $f_{\text{cr}}(p,\mu)$ for the runs with mono-energetic (left) and power-law (right) CRs at different times.
    The mono-energetic CRs isotropize at late times.
    For the power-law case, lower-energy CRs isotropize with time, unlike the higher-energy CRs.
    It is indicated with three white contours: $4 \pi p^2 f_{\text{cr}} \approx  1.6 \times 10^{-3}$ (left), $\ 6.2 \times 10^{-5}$ (middle), $\ 4 \times 10^{-6}$ (right).
    The white dashed line marks $p'_{\rm eff} \approx 9.3 p'_{\min}$ (Equation \ref{eq:p_cut}), a cutoff above which the CRs do not isotropize effectively till saturation ($t \approx 10 \gamma_{\text{fast}}^{-1}$).}
    \label{fig:momangdis_comparison_HI}
\end{figure}

We discuss how the isotropization of mono-energetic CRs differs from the power-law CRs.
We use our benchmark runs MEI:5 and PLI:5-400 (I1 and I4 in Table \ref{tab:sims_data}), and analyze the CR distributions in momenta $(p)$ and direction cosines $(\mu=p_x/p)$ in plasma ion rest frame.
Figure \ref{fig:momangdis_comparison_HI} shows the CR distributions for the two runs.
For MEI:5, the anisotropic distribution at $t \approx 3 \gamma_{\text{fast}}^{-1}$ (more particles have positive $\mu$) becomes mostly isotropic during saturation at $t \approx 10 \gamma_{\text{fast}}^{-1}$.
For PLI:5-400, we observe the anisotropy at $t \approx 3 \gamma_{\text{fast}}^{-1}$ (more particles with a particular momentum have positive $\mu$; three contours of $4 \pi p^2 f_{\text{cr}}$ show the anisotropy).
At $t \approx 10 \gamma_{\text{fast}}^{-1}$, the lower-energy CRs isotropize, and the higher-energy CRs remain unaffected (limiting their contribution to the saturated field; see Section \ref{sec:saturation_vs_CRdistribution}). 
However, later at $t \approx 150 \gamma_{\text{fast}}^{-1}$, the higher-energy CRs slowly isotropize.
We will show in Section \ref{sec:NRSI_by_high_energy_tail} that, in the absence of the lower-energy CRs, the higher-energy ones effectively contribute to the saturated field, and isotropize as well.

We estimate an effective cutoff momentum $p'_{\rm eff}$ for power-law CRs with $p'_{\min}-p'_{\max}$, such that the CRs with momenta $p'_{\min}-p'_{\rm eff}$ isotropize and contribute to the saturated field, but CRs with $p'_{\rm eff}-p'_{\max}$ do not.
Therefore, the saturated field would be $\sqrt{\xi[\text{PL:}p'_{\min}-p'_{\rm eff}]/2}$ (following Equation \ref{eq:xi_vs_saturation}).
Equating this with $\exp(0.4) \times \sqrt{\xi[\text{ME:}p'_{\text{min}}]/2}$ (maximum saturated field for power-law CRs, see the corollary below Equation \ref{eq:xi_vs_saturation_fitting}), we get
\begin{equation} \label{eq:p_cut}
    p'_{\rm eff} \approx \exp(\exp(0.8)) p'_{\text{min}}
    \implies
    p'_{\rm eff} \approx 9.3 p'_{\text{min}}.
\end{equation}
The white dashed lines in Figure \ref{fig:momangdis_comparison_HI} mark $p \approx p'_{\rm eff}$. 
Till saturation (right middle box), CRs with $p \gtrsim p'_{\rm eff}$ do not isotropize effectively.
This cutoff momentum in Equation \ref{eq:p_cut} is only valid for $f'_{\rm cr}(p) \propto p^{-4}$; it decreases for $f'_{\rm cr}(p) \propto p^{-5}$ and increases for $f'_{\rm cr}(p) \propto p^{-3}$ (see Appendix \ref{sec:cr_with_1byp35}).

Using $p'_{\rm eff}$ in Equation \ref{eq:p_cut}, we evaluate the effective anisotropy parameters $\xi_{\rm eff}$ (Equation \ref{eq:xi_effective}) for our runs. 
The lower inset of Figure \ref{fig:saturation_vs_xi} shows the effective scaling of the saturated fields with $\xi_{\rm eff}$ (Equation \ref{eq:saturation_vs_xi_eff}).
This scaling is also valid for nonrelativistic CRs (see Appendix \ref{sec:nonrel_crs}).

\subsubsection{High-energy CRs with $p'_{\rm eff} \leq p' \leq p'_{\max}$ in absence of lower-energy CRs} \label{sec:NRSI_by_high_energy_tail}

We show that, in the absence of the lower-energy CRs, the higher-energy ones with $p' \gtrsim p'_{\rm eff}$ drive NRSI, contribute to its saturated field, and isotropize.
Note that this study is different from Section \ref{sec:modelling_simulation_saturation}, where we varied $p'_{\min}$ or $p'_{\max}$ by keeping $n_{\rm cr}$ fixed, and found that saturation is more sensitive to $p'_{\min}$. 
Unlike that, here we vary both $n_{\rm cr}$ and $p'_{\min}$, essentially studying NRSI by higher-energy CRs in absence of lower-energy ones.

\begin{figure}
    \centering
    \subfloat[Growth and saturation \label{fig:MagFld_diff_ncr_a}]{
        \includegraphics[width=0.47\textwidth]{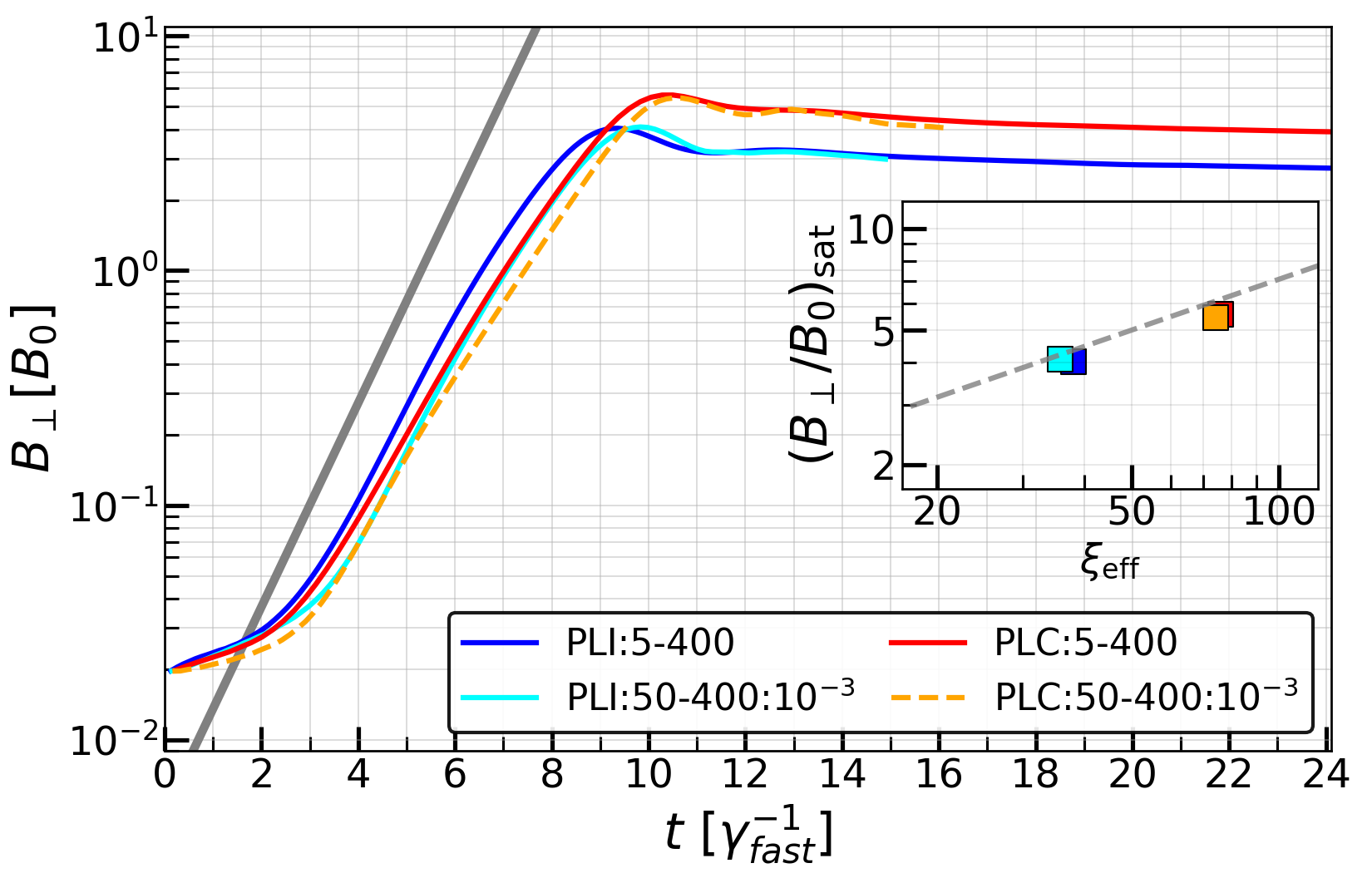}
    }
    \hfill
    \subfloat[CR distribution \label{fig:MagFld_diff_ncr_b}]{
        \includegraphics[width=0.47\textwidth]{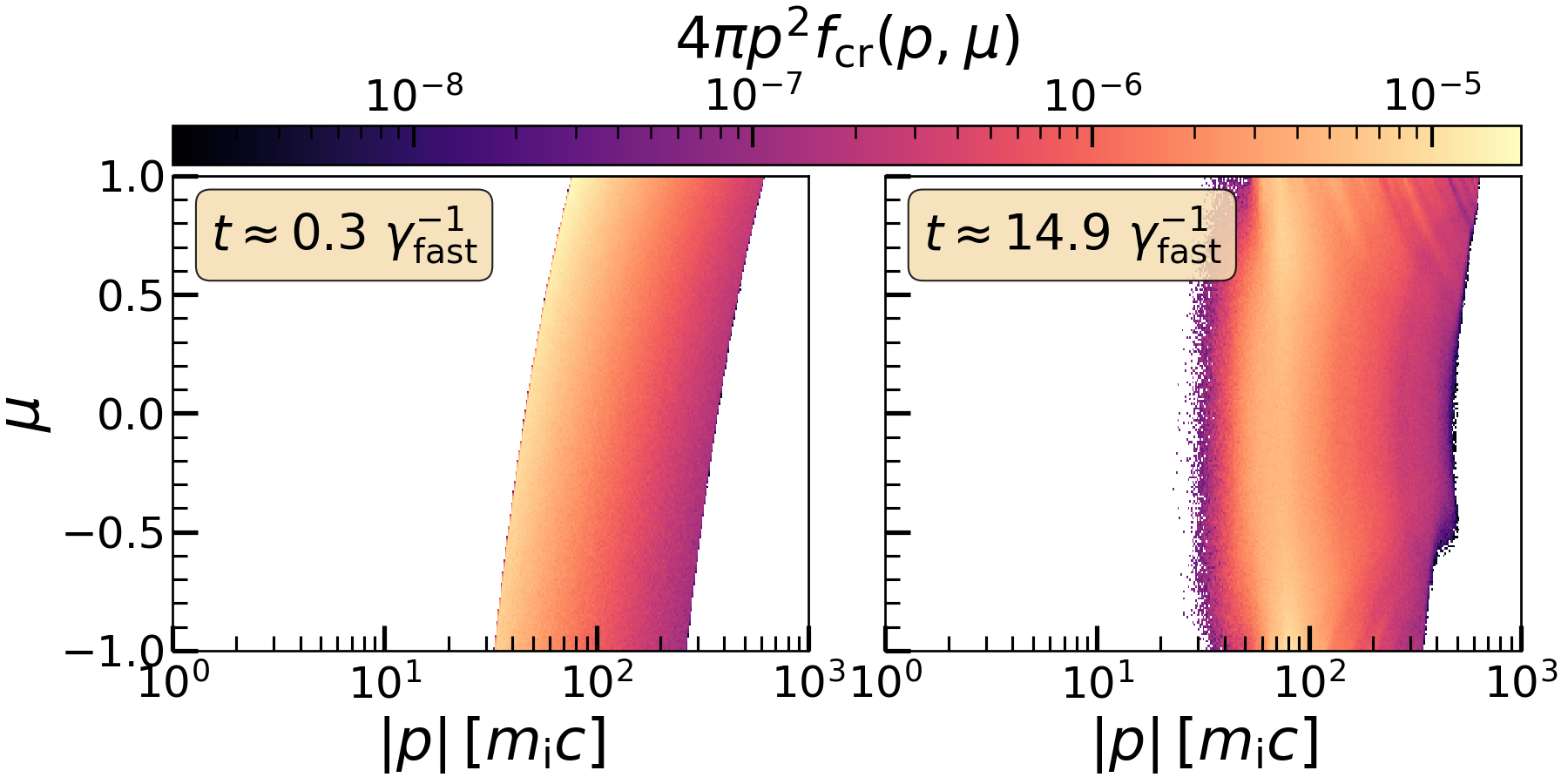}
    }
    \caption{[Top panel] Transverse magnetic fields $B_\perp$ for runs I4, C4, I10, and C10 in Table \ref{tab:sims_data}.
    Inset: $(B_{\perp}/B_0)_{\rm sat}$ versus $\xi_{\rm eff}$, modeled with Equation \ref{eq:saturation_vs_xi_eff} (grey dashed line) for the runs with squares of respective colors.
    [Bottom panel] CR distributions for PLI:50-400:$10^{-3}$. 
    During saturation, roughly all the higher-energy CRs isotropize, efficiently contributing to the saturated field (in absence of the lower-energy ones).}
    \label{fig:MagFld_diff_ncr}
\end{figure}

We consider two power-law CR distributions: (I) the full spectrum with $p'_{\min} \leq p' \leq p'_{\max}$ and number density $n_{\rm cr} [p'_{\min}-p'_{\max}]$, (II) the higher-energy CRs with $p'_{\rm eff} \leq p' \leq p'_{\max}$ and number density $n_{\rm cr} [p'_{\rm eff}-p'_{\max}] \sim 0.1 \times n_{\rm cr} [p'_{\min}-p'_{\max}]$ (as $p'_{\rm eff} \sim 10 p'_{\min}$, see Equation \ref{eq:p_cut}).
For our benchmark runs PLI:5-400 and PLC:5-400 (I4 and C4 in Table \ref{tab:sims_data}), $p'_{\min} \sim 5 m_i c$, $p'_{\max} \sim 400 m_i c$, and $n_{\rm cr}/n_0 = 10^{-2}$.
The higher-energy tails for these runs will have $p'_{\min} = p'_{\rm eff} \sim 50m_i c$, $p'_{\max} \sim 400 m_i c$, and $n_{\rm cr}/n_0 \sim 10^{-3}$.

The simulations with these higher-energy CRs are PLI:50-400:$10^{-3}$ and PLC:50-400:$10^{-3}$ (I10 and C10 in Table \ref{tab:sims_data}).
The top panel of Figure \ref{fig:MagFld_diff_ncr} shows the growth of $B_{\perp}$ for these runs.
Although they look similar, the growth for PLI:50-400:$10^{-3}$ and PLC:50-400:$10^{-3}$ are slower than PLI:5-400 and PLC:5-400, respectively (compare growth rates from Table \ref{tab:sims_data}).
The saturated magnetic fields for the higher-energy CRs and the whole spectra are shown in the inset; they scale with $\xi_{\rm eff}$ following Equation \ref{eq:saturation_vs_xi_eff}.
The bottom panel of Figure \ref{fig:MagFld_diff_ncr} shows that CR distributions for PLI:50-400:$10^{-3}$ in plasma-ion rest frame isotropize effectively after saturation.
Thus, in the absence of the lower-energy CRs, the higher-energy CRs contribute to the saturated field.

\section{Astrophysical Implications} \label{sec:discussion}

\begin{figure*}
    \centering
    \includegraphics[width=0.9\textwidth]{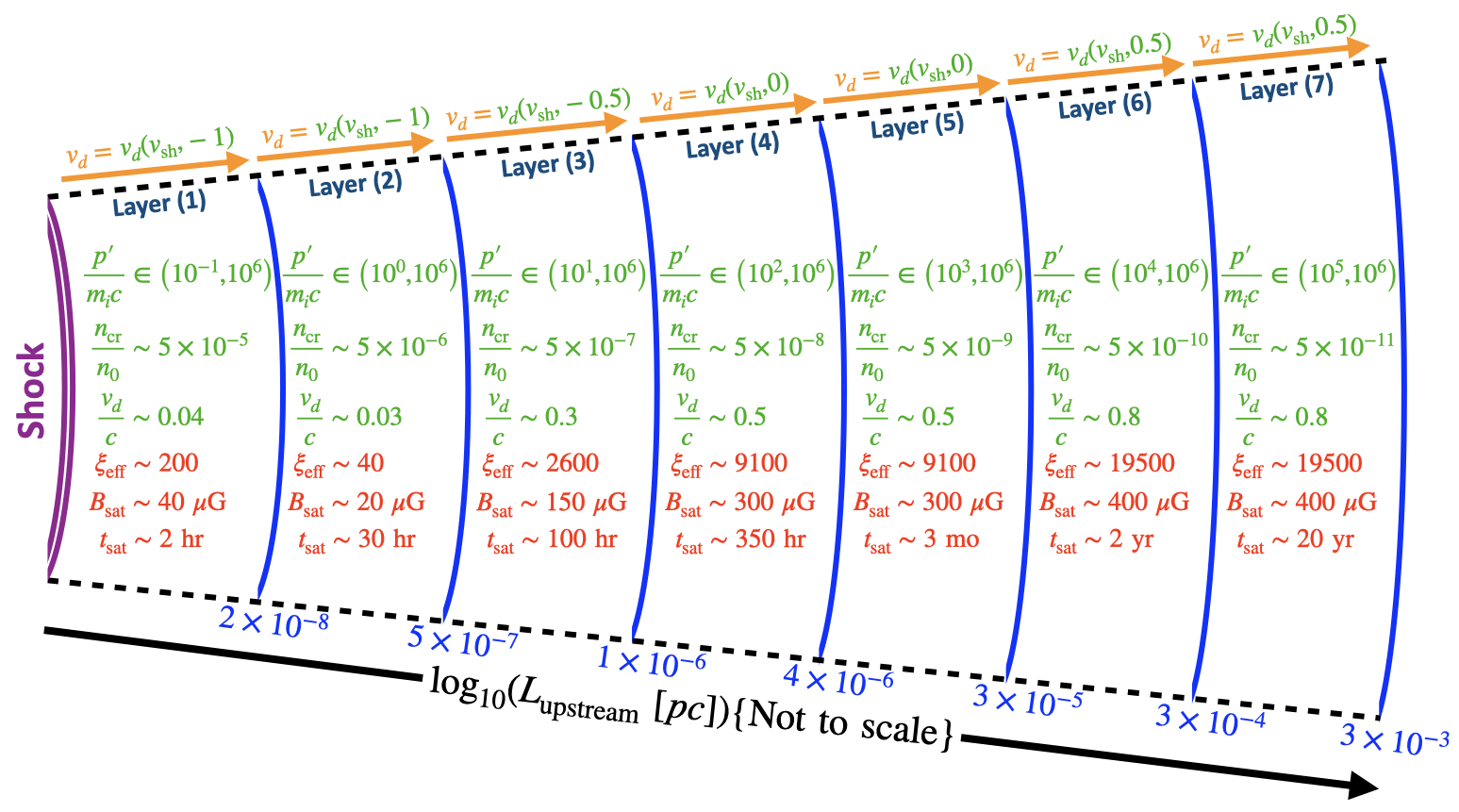}
    \caption{Schematic illustration of CR self-confinement via NRSI in a layered upstream ISM plasma ahead of an SNR shock.
    CRs with $f'_{\rm cr} (p) \propto p^{-4}$, produced at the shock, enter an initially unperturbed upstream region and amplify magnetic fluctuations through NRSI.
    CRs with momenta $p' \in (p'^{\ (i)}_{\min}, 10p'^{\ (i)}_{\min})$ isotropize in each layer, where $p'^{\ (i)}_{\min}$ is the minimum momentum in layer $(i)$ (as $p'^{\ (i)}_{\rm eff} \sim 10 p'^{\ (i)}_{\min}$ in Equation \ref{eq:p_cut}; see Section \ref{sec:simulation_isotropization} for details).
    Higher-energy particles escape to the next unperturbed region, where they become fresh drivers of NRSI with $p'^{\ (i+1)}_{\min}$ increased by a factor of 10 and reduced number density $n^{(i+1)}_{\rm cr}$ compared to layer $(i)$.
    The CR drift speed $v^{(i)}_d = v_d (v_b,\mu'^{\ (i)}_{\min})$ depends on the boost speed $v_b$ and minimum pitch angle $\mu'^{\ (i)}_{\min}$ (see Equation \ref{eq:drift_speed}).
    In all layers, $v_b$ equals the shock speed $v_{\rm sh}$ (transforming to upstream plasma rest frame). 
    In farther layers, we assume that $\mu'^{\ (i)}_{\min}$ increases, implying more forward-beamed distributions.
    After the instability saturates at time $t^{(i)}_{\rm sat} \sim 10/ \gamma_{\rm fast}^{(i)}$ in layer $(i)$, CRs of higher momenta diffuse to the next layer over a distance comparable to their Larmor radii, i.e. $\Delta L^{(i)}_{\rm upstream} \sim 10 p'^{\ (i)}_{\min}c/e B^{(i)}_{\rm sat}$.}
    \label{fig:SNR_LayerISM_cartoon}
\end{figure*}

In astrophysical shocks (e.g., SNRs) or the ISM plasmas, the CR distributions can deviate substantially from commonly assumed idealized forms, such as isotropic mono-energetic or power-law ($f'_{\rm cr} (p) \propto p^{-4}$) populations.
The linear growth of current-driven NRSI is independent of the CR distribution (see Section \ref{sec:simulation_linear_growth}).
Therefore, any CR spectrum drifting through ISM plasma satisfying Bell's criterion $r_{L,\min}^{\prime -1} <k_{\rm fast} < d_i^{-1}$ drives NRSI, making this instability effective for the magnetic field amplification and scattering/self-confinement of CRs.

However, as expected from DSA, CRs naturally follow a power-law momentum distribution with $f'_{\rm cr} (p) \propto p^{-4}$. 
These CRs are self-confined in the upstream plasma through isotropization via NRSI in the plasma-rest frame.
In the shock rest frame, the CRs, along with the upstream plasma,  are carried back to the shock front and further injected into DSA, reaching PeV energies.

For power-law distributions with $p'_{\min} \leq p' \leq p'_{\max}$ and $p'_{\max} \gg p'_{\min}$, the CRs with momentum $p' \lesssim 10 p'_{\min}$ isotropize and contribute to the saturation of NRSI (Section \ref{sec:simulation_isotropization}). 
The higher-energy CRs, in the presence of the lower-energy ones, do not isotropize effectively, and may escape a plasma region before getting confined, potentially hindering their acceleration to higher energies.

Our results suggest a layered picture of the upstream plasma for CR confinement and magnetic field amplification.
Lower-energy CRs are self-confined closer to the shock.
Higher-energy CRs, leaking further ahead, enter an unperturbed plasma region and act as fresh drivers of NRSI.
In the absence of the lower-energy CRs, the higher-energy ones can drive NRSI, contribute to its saturation, and isotropize (Section \ref{sec:NRSI_by_high_energy_tail}).
Thus, the power-law CRs amplify magnetic fields over successive layers; each decade of CR momenta gets confined in each layer.

There are two limitations of this model. 
Firstly, the higher-energy CRs in the successive upstream layers have a smaller number density, leading to significantly slower growth of NRSI.
Secondly, the saturated magnetic fields in farther layers are small. 
To resolve these issues, we use different pitch-angle distributions of CRs in the upstream layers. 
We implement more conical distributions for higher-energy CRs in farther layers. 
This is reasonable for more anisotropic high-energy CRs away from the shock. 
This anisotropy produces a larger CR drift speed $v_d$ and a larger mean momentum $p_d$. 
Thus, the growth is faster, and the saturated field is larger.

We estimate the saturated magnetic fields, saturation timescales, and CR confinement lengthscales in each layer of our schematic model using typical ISM plasma values. 
Figure \ref{fig:SNR_LayerISM_cartoon} shows an illustration of this model and the estimates. 
The background magnetic field $B_0$, directed radially outward, corresponds to an Alfv\'en speed
\begin{equation}
    v_{A0} \approx 2 \times 10^{-5} c
    \left[ \frac{B_0}{3 {\rm \mu G}} \right]
    \left[ \frac{n_0}{1{\rm cm}^{-3}} \right]^{-1/2},
\end{equation}
where $n_0$ is the ISM number density.

We consider CRs, drifting along the background magnetic field, with $p'_{\min}\sim 10^{-1} m_i c$, $p'_{\max} \sim 10^6 m_ic \sim {\rm PeV}/c$, and $f'_{\rm cr} (p) \propto p^{-4}$. 
Here, the minimum momentum $p'_{\min}$ corresponds to the CR energy $10E_{\rm sh}$, i.e. $p'_{\min} \sim \sqrt{20 m_i E_{\rm sh}}$, where $E_{\rm sh}=0.5 m_i v_{\rm sh}^2$ \citep[see e.g.][]{Caprioli_2014} and $v_{\rm sh} \approx 0.04 c$ is the shock speed. 
We take the CR acceleration efficiency 
\begin{equation}
    \eta = \frac{U_{\rm cr}}{U_{\rm sh}} 
    = \frac{c \int_{p'_{\min}}^{p'_{\max}} p f'_{\rm cr} d^3 p}{0.5 m_i n_0 v_{\rm sh}^2} 
\end{equation}
to be $10 \%$, where $U_{\rm cr}$ and $U_{\rm sh}$ are CR energy density\footnote{Here, we use the relativistic relation $E(p)=pc$ even for the lowest-energy CRs, which are nonrelativistic. The error introduced by this is small.} and shock kinetic energy density, respectively. 
Thus, the CR number fraction is given by
\begin{equation}
\begin{split}
    \frac{n_{\rm cr}}{n_0} \approx 5 \times 10^{-5} 
    &\left[ \frac{\eta}{0.1} \right]
    \left[ \frac{v_{\rm sh}}{0.04 c} \right]^2 \times \\
    &\left[ \frac{p'_{\min}}{10^{-1}m_i c} \right]^{-1}
    \left[ \frac{\ln \left( \frac{p'_{\max}}{p'_{\min}} \right)}{ \ln \left( 10^7 \right) } \right]^{-1}.
\end{split}
\end{equation}
CRs with the first decade of momenta isotropize in each layer, and the rest of them drift to the subsequent layers.
Thus, the CR number density in subsequent layers decreases by a factor of 10; for layer $(i)$, $n^{(i)}_{\rm cr} \propto 1/p'^{\ (i)}_{\min}$.

We consider the minimum pitch angle $\mu'^{\ (i)}_{\min}$ in layer $(i)$ as a free parameter. 
For the first two layers close to the shock, $\mu'_{\min}=-1$, i.e. CRs are isotropic (see the discussion below Equation \ref{eq:distribution_PL}). 
In the successive layers, $\mu'_{\min}$ gradually increases, i.e. the distribution becomes more conical.
The drift speed $v^{(i)}_d$, given by the functional form as $v^{(i)}_d = v_d ( v_b, \mu'^{\ (i)}_{\min} )$ is calculated from Equation \ref{eq:drift_speed}, with $v_b = v_{\rm sh}$.
The values of $\mu'^{\ (i)}_{\min}$ are indicated at the top of the orange arrows.
How exactly $\mu'^{\ (i)}_{\min}$ evolves with distance from the shock, however, remains an assumption in our model.

We calculate the effective anisotropy parameter $\xi^{(i)}_{\rm eff}$ for the CRs with the power-law distribution using Equation \ref{eq:xi_effective} with $p'^{\ (i)}_{\rm eff} \approx 10 p'^{\ (i)}_{\min}$.
For this, we evaluate the mean momentum $p^{(i)}_d$ using Equation \ref{eq:mean_momentum}.
Following Equation \ref{eq:saturation_vs_xi_eff}, the saturated magnetic field is 
\begin{equation}
    B^{(i)}_{\perp,{\rm sat}} \approx 
    40 {\rm \mu G}
    \left[ \frac{B_0}{3 {\rm \mu G}} \right]
    \left[ \frac{\xi^{(i)}_{\rm eff}}{200} \right]^{1/2},
\end{equation}
and the net amplified magnetic field is $B^{(i)}_{\rm sat} \approx B^{(i)}_{\perp,{\rm sat}}$, as $B^{(i)}_{\perp,{\rm sat}} \gg B_0$.

Using Equation \ref{eq:k_fast_fluid}, the saturation timescale $t^{(i)}_{\rm sat} \approx 10/ \gamma_{\rm fast}^{(i)}$ is 
\begin{equation}
\begin{split}
    t^{(i)}_{\rm sat} \approx 2\ {\rm hr}
    &\left[ \frac{n^{(i)}_{\rm cr}/n_0}{5 \times 10^{-5}} \right]^{-1} \times \\
    &\left[ \frac{v^{(i)}_d}{v_d \left( v_{\rm sh}, -1 \right)} \right]^{-1}
    \left[ \frac{n_0}{1{\rm cm}^{-3}} \right]^{-1/2}.
\end{split}
\end{equation}
The confinement length, which is also the approximate size of layer $(i)$, is the length up to which the CRs with momenta $p'^{\ (i)}_{\min} - 10p'^{\ (i)}_{\min}$ are confined. 
This is calculated as $\Delta L^{(i)}_{\rm upstream} \approx 10 p'^{\ (i)}_{\min}c/e B^{(i)}_{\rm sat}$, i.e.
\begin{equation}
    \Delta L^{(i)}_{\rm upstream} \approx 2 \times 10^{-8} {\rm pc}
    \left[ \frac{p'^{\ (i)}_{\min}}{10^{-1} m_i c} \right]
    \left[ \frac{B^{(i)}_{\rm sat}}{40 {\rm \mu G}} \right]^{-1}.
\end{equation}
We perform the calculations for the subsequent layers accordingly.

Our model and the estimates in Figure \ref{fig:SNR_LayerISM_cartoon} show that CRs up to PeV energies can be confined within a lengthscale of $L_{\rm upstream} \sim 3 \times 10^{-3} {\rm pc}$, much smaller than the SNR itself (e.g. radius $R_{\rm SNR} \sim 0.1 {\rm pc}$).
The confinement timescale (similar to the saturation timescale) is $\sim 20 {\rm yr}$, which is shorter than the free-expansion phase ($\sim 100 {\rm yr}$) and the Sedov-Taylor phase ($\sim 1000 {\rm yr}$) of an SNR.
The magnetic field amplification up to $\sim 400 {\rm \mu G}$, significantly larger than the ISM magnetic field ($\sim 3 {\rm \mu G}$), is also possible.
The short confinement timescale $(\sim 20 {\rm yr})$ of high-energy CRs up to PeV energies implies the feasibility of their production at an early stage of an SNR evolution \citep[e.g.][]{cao2026ultra}.

Although motivated by our results, the model is simplistic and speculative. 
Testing this scenario requires more realistic setups, allowing for particle escape and re-seeding of instabilities in fresh plasma. 
Such models are essential for assessing whether SNR shocks can truly sustain the conditions required for PeV acceleration and account for the observed Galactic CR spectrum up to the `knee' and beyond.

\section{Conclusions} \label{sec:conclusion}
In this work, we have used different distributions of cosmic rays (CRs) to study the linear growth and nonlinear saturation of the nonresonant streaming instability (NRSI) using 1D particle-in-cell simulations.
We show that, during saturation, the CRs with a power-law distribution, having a wide range of energies, behave differently from the CRs with a mono-energetic distribution.
The following is a summary of our key results:
\begin{enumerate}
    \item The linear growth of NRSI depends on the net CR current through the background plasma; the fastest-growing modes and their growth rates are nearly the same across different CR distributions.
    
    \item Saturation of NRSI occurs through the isotropization of CRs in the background plasma, which quenches the CR current.
    Thus, the anisotropy pressure of CRs (difference between parallel and transverse CR pressure as measured in the plasma rest frame) gets transferred to the transverse magnetic pressure and the background plasma (resulting in plasma heating).
    
    \item Saturation of NRSI depends on the CR distributions.
    Mono-energetic CRs contribute to the saturation and isotropize.
    In contrast, for the power-law CRs, only the lowest-energy particles contribute to the saturation; the isotropization of the highest-energy CRs is inefficient, limiting their contribution to the saturated field.
    However, in the absence of the low-energy CRs, the high-energy ones effectively amplify the transverse fields and isotropize. 
    
    \item We tested the effectiveness of the anisotropy parameter $\xi$, the ratio of CR anisotropy pressure to the magnetic pressure (Equation \ref{eq:xi_gupta_2021}), which predicts the saturated magnetic field (Equation \ref{eq:xi_vs_saturation}). 
    For mono-energetic CRs, $\xi$ accurately estimates the saturated field, consistent with the previous findings \citep{gupta_2021,zacharegkas2024modeling}.
    However, for power-law CRs extending across a wide momentum range, the usual $\xi$ overestimates the saturated field, as it does not account for the negligible contribution of the high-energy CRs.
    
    \item We introduce an effective anisotropy parameter $\xi_{\rm eff}$ (Equation \ref{eq:xi_effective}), which predicts the  saturated fields for power-law CRs (Equation \ref{eq:saturation_vs_xi_eff}).
    It accounts for the fact that only the low-energy CRs with $p' \lesssim p'_{\rm eff}$ contribute to the saturation; $p'_{\rm eff} \sim 10p'_{\min}$ for $f'_{\rm cr} \propto p^{-4}$ (Equation \ref{eq:p_cut}).
    Our prediction holds for both relativistic and nonrelativistic CRs.

\end{enumerate}

Based on these findings, we propose a plausible model for CR confinement via NRSI in the ISM plasma upstream of an SNR shock (see Figure \ref{fig:SNR_LayerISM_cartoon}).
Power-law CRs enter the unperturbed ISM and excite NRSI.
The lowest-energy CRs get confined closer to the shock; the higher-energy CRs escape and enter the subsequent unperturbed ISM.
Thus, the magnetic field is amplified and the CRs are confined over successive layers in the upstream ISM. 
We estimate the amplified magnetic fields and the length and timescales of CR confinement, suitable for an SNR.
Future realistic simulations will be crucial to test the viability of this distribution-dependent confinement in astrophysical environments.


\begin{acknowledgments}
We thank the anonymous referee for insightful comments and constructive suggestions.
S.D. thanks Tanu Sharma and Rahul Verma for fruitful discussions.
S.G. thanks Damiano Caprioli and Anatoly Spitkovsky for stimulating discussions on the cosmic ray streaming instabilities. S.G. acknowledges financial support for his postdoctoral research at Princeton University through NSF grant PHY-2206607 and Simons Foundation grant MP-SCMPS-00001470.
\end{acknowledgments}

%

\vspace{5mm}
\facilities{Param-Pravega Supercomputing Facility (National Supercomputing Mission, Indian Institute of Science)}





\appendix

\section{Kinetic Dispersion of NRSI} \label{sec:derivation_kinetic_dispersion}

We derive the kinetic dispersions similar to \citet{amato2009}.
By linearizing the Vlasov equation, we get the dispersion relation for parallel-propagating transverse waves in a composite plasma, given by 
\begin{align}
    \frac{k^2 c^2}{\omega^2}
    &= 1 
    + \sum_{s} \frac{4\pi^2 q_s^2}{\omega}
    \int_0^\infty dp \int_{-1}^1 d\mu\;
    \frac{p^2 v(p)\left(1-\mu^2\right)}
         {\omega - k v(p)\mu - \Omega_s}
    \nonumber \\
    &\quad \times \left[
        \frac{\partial f_s}{\partial p}
        + \left( \frac{k v(p)}{\omega} - \mu \right)
        \frac{1}{p}
        \frac{\partial f_s}{\partial \mu}
    \right],
    \label{eq:dispersion_general_kinetic}
\end{align}
where $q_s$ and $\Omega_s=\dfrac{q_sB_0}{\gamma m_s c}$ are the charge and relativistic gyrofrequency of species $s$, respectively.
Replacing $f_s$ with the distributions of cold ions and electrons, and CRs, we evaluate the imaginary solutions of $\omega$, which provide the growth rates ($\gamma$) of RSI and NRSI.

The growth rates of NRSI for isotropic (I) CRs are 
\begin{equation}
\gamma_{\rm I}(k)=
    \frac{1}{\sqrt{2}} \sqrt{-\Theta_{\rm I}(k)
    + \sqrt{ \Theta_{\rm I}(k)^2 + \big(\alpha(k) Y_{\rm I}(k)\big)^2}},
\label{eq:dispersion_iso}
\end{equation}
where $\Theta_{\rm I}(k) = k^2 v_{A0}^2 + \alpha(k)\left(1 + X_{\rm I}(k)\right)$ and $\alpha (k) = -\dfrac{n_{\text{cr}}}{n_0} k v_d \omega_{ci}$. 
For mono-energetic (ME) CRs, $X_{\rm I} (k)$ and $Y_{\rm I} (k)$ are 
\begin{subequations} 
    \begin{align}
        X_{\rm MEI}(k) &= - \frac{1}{2} \frac{p'_0}{( kr'_{L,0} )^2} \ln \left| \frac{1+kr'_{L,0}}{1-kr'_{L,0}} \right| \\
        Y_{\rm MEI}(k) &= -\frac{\pi}{2} \frac{p'_0}{( kr'_{L,0} )^2} \times 
        \begin{dcases}
            0,\ \left(kr'_{L,0}<1\right) \\
            1,\ \left(kr'_{L,0} \geq 1\right)
        \end{dcases}
        .
    \end{align}
\end{subequations}
For power-law (PL) CRs, $X_{\rm I} (k)$ and $Y_{\rm I} (k)$ are
\begin{subequations}
\begin{align}
X_{\rm PLI}(k)
&= \frac{\mathcal{A}}{4}
\Bigg[
  \left( \frac{1}{s^4} - \frac{1}{s^2} \right)
  (s^2 - 1)\,
  \ln\left|\frac{1+s}{1-s}\right|
\notag \\
&\qquad\qquad
  + 2\left( \frac{1}{s^3} + \frac{1}{s} \right)
\Bigg]_{kr'_{L,\min}}^{kr'_{L,\max}}
\\[4pt]
Y_{\rm PLI}(k)
&= \frac{\pi \mathcal{A}}{4}\, 
\left[
  \frac{2}{s^2} - \frac{1}{s^4}
\right]_{{\rm Max}\left[1,\, kr'_{L,\min}\right]}^{kr'_{L,\max}}
\end{align}
\end{subequations}
where $\mathcal{A=}\left( \dfrac{1}{kr'_{L,\min}} - \dfrac{1}{kr'_{L,\max}} \right)^{-1}$ and $r'_L = \dfrac{p'c}{eB_0}$ is the Larmor radius.

Comparing the expressions of $X_{\rm PLI}(k)$ and $Y_{\rm PLI}(k)$ with those of $I_1^+ (k)$ and $I_2 (k)$ (their Equations 24 and 25) in \citet{amato2009}, we observe that the terms with $1/s^3$ and $1/s^4$ are absent.
This causes the growth rates of RSI and NRSI to be almost similar for $k<1/r'_{L,\min}$ (see their Figure 1).
However, these terms cannot be neglected for $k < 1/r'_{L,\min}$, as they cause the growth rates of RSI to be larger than NRSI; this has important consequences for CR scattering. 

The growth rates of NRSI for cone (C) CRs are
\begin{equation} \label{eq:dispersion_cone}
    \gamma_{\rm C}(k) = \frac{1}{2} \left[ \Phi (k) + \Im \left\{ \sqrt{ \Theta_{{\rm C}1}(k)^2 + 4 \Theta_{{\rm C}2}(k) } \right\} \right],
\end{equation}
where $\Theta_{{\rm C}1}(k) = \dfrac{n_{\text{cr}}}{n_0} \omega_{ci} \left( 1 - X_{\rm C}(k) \right) + j \Phi(k)$, $\Theta_{{\rm C}2}(k) = \alpha (k) + k^2 v_{A0}^2$, $\Phi(k) = \dfrac{n_{\rm cr}}{n_0} \omega_{ci} Y_{\rm C} (k)$.
For mono-energetic (ME) CRs, $X_{\rm C} (k)$ and $Y_{\rm C} (k)$ are
\begin{subequations} 
    \begin{align}
        X_{\rm MEC}(k) &= \frac{p'_0}{( kr'_{L,0} )^2} \ln \left| 1 - kr'_{L,0} \right| \\
        Y_{\rm MEC}(k) &= - \frac{\pi p'_0}{( kr'_{L,0} )^2} \times 
        \begin{dcases}
            0 \text{ ; if } kr'_{L,0}<1 \\
            1 \text{ ; if } kr'_{L,0} \geq 1
        \end{dcases}
        .
    \end{align}
\end{subequations}
For power-law (PL) CRs, $X_{\rm C} (k)$ and $Y_{\rm C} (k)$ are
\begin{subequations}
\begin{align}
X_{\rm PLC}(k)
&= \frac{\mathcal{A}}{2}
\Bigg[
  \left( 1 - \frac{2}{s^2} + \frac{1}{s^4} \right)
  \ln\left|1 - s\right|
\notag \\
&- \ln\left|s\right|
  + \left( \frac{1}{s} + \frac{1}{2s^2} + \frac{1}{s^3} \right)
\Bigg]_{k r'_{L,\min}}^{k r'_{L,\max}}
\\[4pt]
Y_{\rm PLC}(k)
&= \frac{\pi \, \mathcal{A}}{2}
\left[
  \frac{2}{s^2} - \frac{1}{s^4}
\right]_{{\rm Max}\left[1,k r'_{L,\min}\right]}^{k r'_{L,\max}}.
\end{align}
\end{subequations}

\section{Helicity of unstable modes} \label{sec:Helicity}

We calculate the helicity $\Delta \phi (k)$ using the formula
\begin{equation} \label{eq:helicity}
    \Delta \phi (k) = \tan^{-1} \left[ \frac{V(k)}{\left\{ Q^2(k) + U^2(k) \right\}^{1/2}}  \right]. 
\end{equation}
We calculate the Stokes parameters $Q(k)$, $U(k)$, and $V(k)$, using the Fourier transforms of the transverse magnetic field components, as
\begin{subequations} \label{eq:stokes_from_fourier}
    \begin{align}
        Q(k) &= \Re \left( \left[ \tilde{B}_y(k) \tilde{B}^*_y(k) - \tilde{B}_z(k) \tilde{B}^*_z(k) \right] \right)\\
        U(k) &= \Re \left( \left[ \tilde{B}_y(k) \tilde{B}^*_z(k) + \tilde{B}_z(k) \tilde{B}^*_y(k) \right] \right)\\
        V(k) &= \Re \left( \left[ \tilde{B}_y(k) \tilde{B}^*_z(k) - \tilde{B}_z(k) \tilde{B}^*_y(k) \right]/j \right).
    \end{align}
\end{subequations}
Here, $\tilde{B}_i (k)$ is the Fourier transform of $B_i(x)$ along the $x$-direction and $\tilde{B}^*_i (k)$ is its complex conjugate.

\section{NRSI Growth Rates from Simulation} \label{sec:NRSI_Growth_Rates_Sims}

\begin{figure}
    \centering
    \subfloat[Growth of a few R-handed modes \label{fig:MagFld_modes_time_evol_a}]{
        \includegraphics[width=0.47\textwidth]{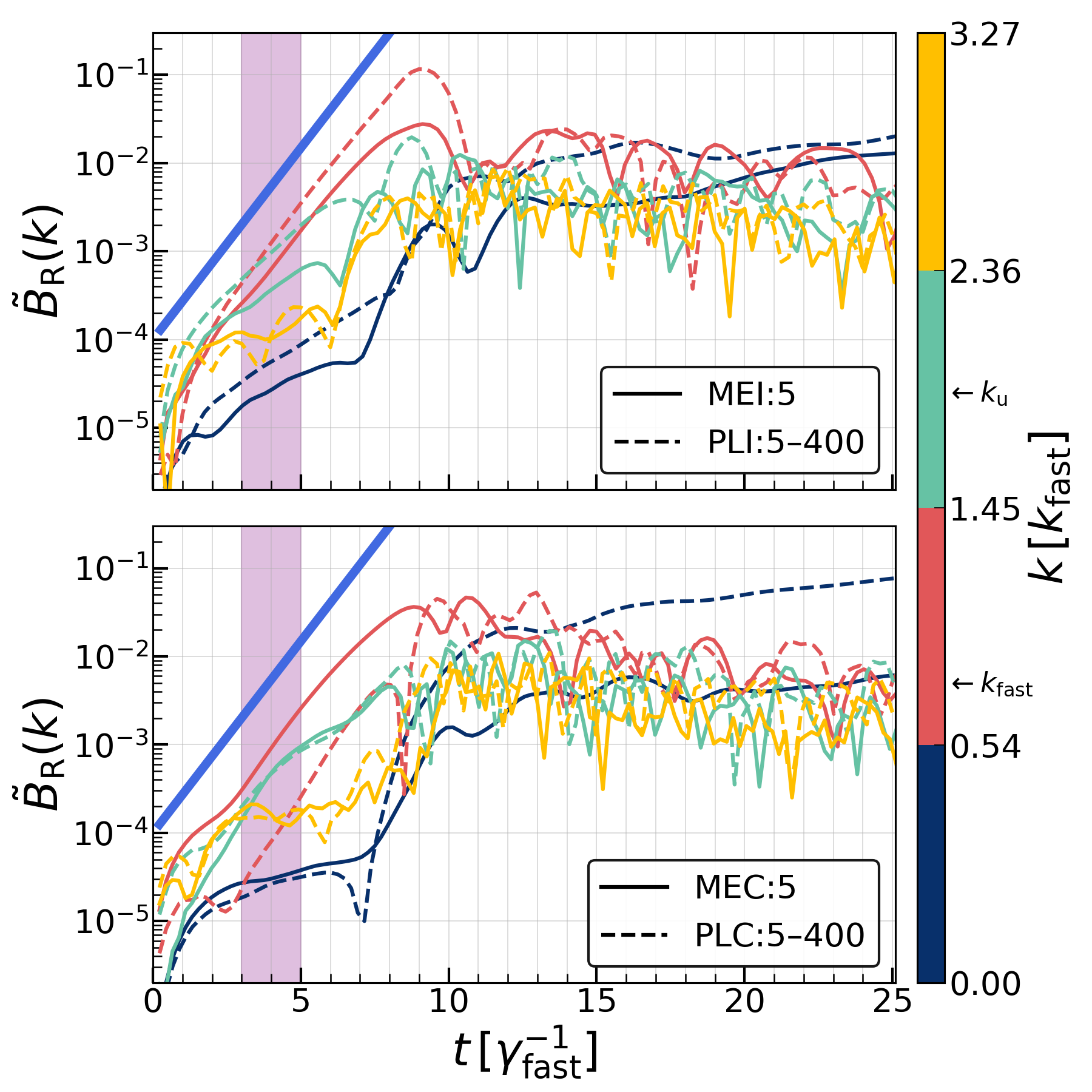}
    }
    \hfill
    \subfloat[Growth Rates of all R-handed modes \label{fig:MagFld_modes_time_evol_b}]{
        \includegraphics[width=0.47\textwidth]{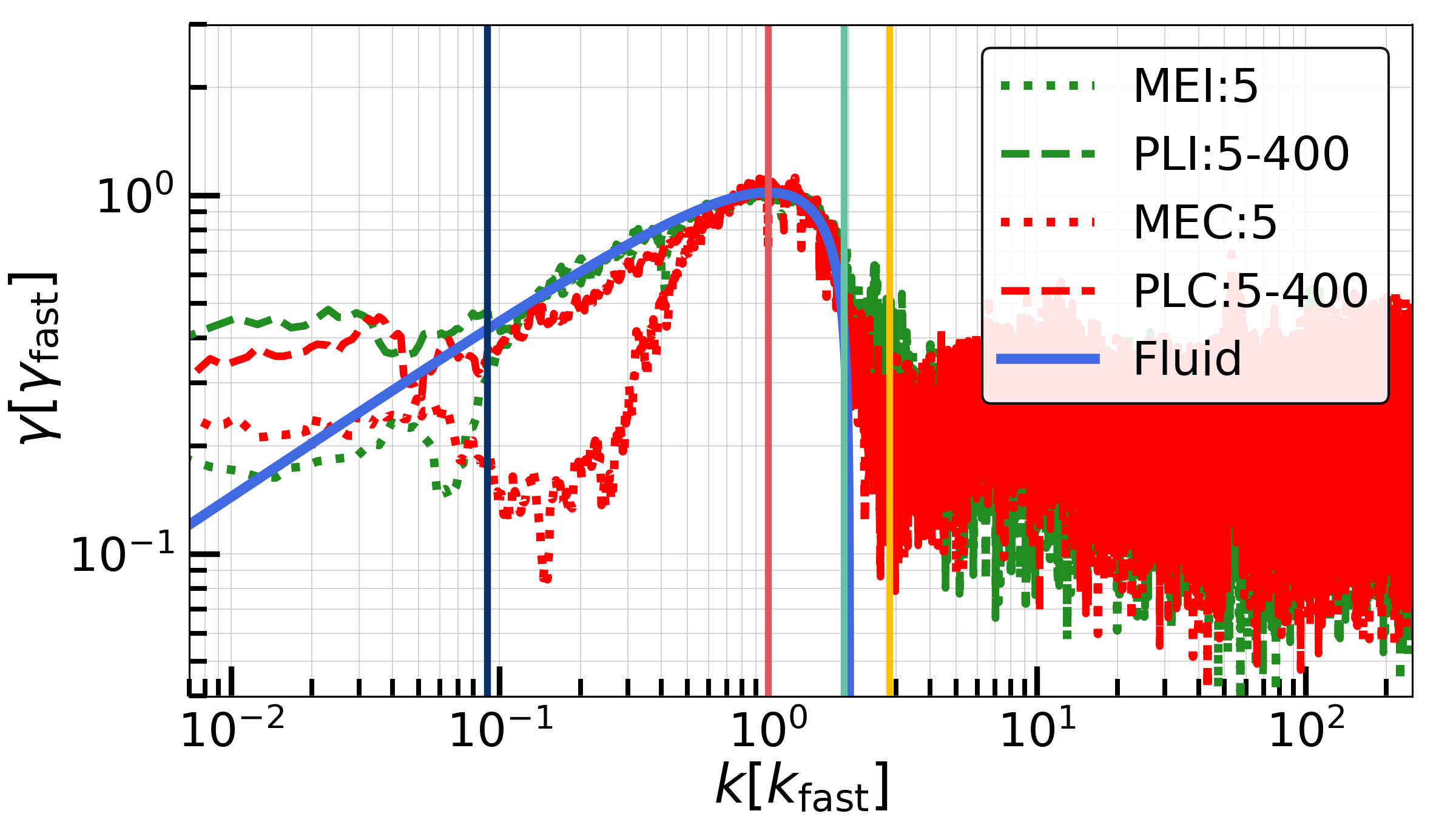}
    }
    \caption{[Top panel] Evolution of a few R-handed ($\tilde{B}_{\text{R}} = | \tilde{B}_y (k) + j \tilde{B}_z (k) |$) components of transverse magnetic fields in the benchmark isotropic (upper panel) and cone (lower panel) runs.
    The color palette in the colorbar represents different modes.
    The growth of the modes $k \sim k_{\text{fast}}$ matches the analytical expectation (blue solid lines).
    We calculate their growth rates numerically within $t \in [3,5] \gamma_{\rm fast}^{-1}$ shaded in purple. 
    [Bottom panel] Growth rates of all the R-handed modes in the benchmark runs.
    The dominant modes match the analytical dispersion in Equation \ref{eq:growth_rate_fluid} (blue solid line).
    The vertical lines mark the modes which we plot in the top panel. }
    \label{fig:MagFld_modes_time_evol}
\end{figure}

We use the Fourier transforms, $\tilde{B}_y$ and $\tilde{B}_z$, and calculate the amplitudes of the R-handed components $\tilde{B}_{\text{R}}$, using the expression 
\begin{equation}
    \tilde{B}_{\rm R} (k) = | \tilde{B}_y (k) + j \tilde{B}_z (k) |.
\end{equation}
The growth of $\tilde{B}_{\text{R}}$ signifies NRSI.
Figure \ref{fig:MagFld_modes_time_evol_a} shows the evolution of $\tilde{B}_{\rm R}$ for a few modes, plotted with different colors, given in the colorbar.
For all the distributions, the exponential growth of $\tilde{B}_{\text{R}}$ for the modes $k \sim k_{\text{fast}}$ matches the analytical prediction, i.e. Equation \ref{eq:k_fast_fluid}, indicated by the blue solid lines. 
The rest of the subdominant modes grow with a slower growth rate.

We evaluate the growth rates of all the R-handed components by numerically fitting their amplitudes within the time range $t \in [3,5] \gamma_{\rm fast}^{-1}$. 
Figure \ref{fig:MagFld_modes_time_evol_b} shows the growth rates for the benchmark runs. 
The growth rates of the dominant modes near $k \sim k_{\rm fast}$ match the analytical dispersion in Equation \ref{eq:growth_rate_fluid}, plotted by the blue solid line.
In fact, the growth rates qualitatively resemble the kinetic dispersions in Figure \ref{fig:dispersion_analytical}. 
However, the deviations for the subdominant modes occur due the difficulty in evaluating their growth rates. 
As we see in Figure \ref{fig:MagFld_modes_time_evol_a}, it is not trivial to select a common time interval within which all the modes grow monotonically as per the analytical expectation.
This leads to erroneous fitting and inaccurate evaluation of the growth rates for the slowly-growing modes.

\section{Different power-law indices} \label{sec:cr_with_1byp35}

\begin{figure}
    
    \centering
    \includegraphics[width=0.47\textwidth]{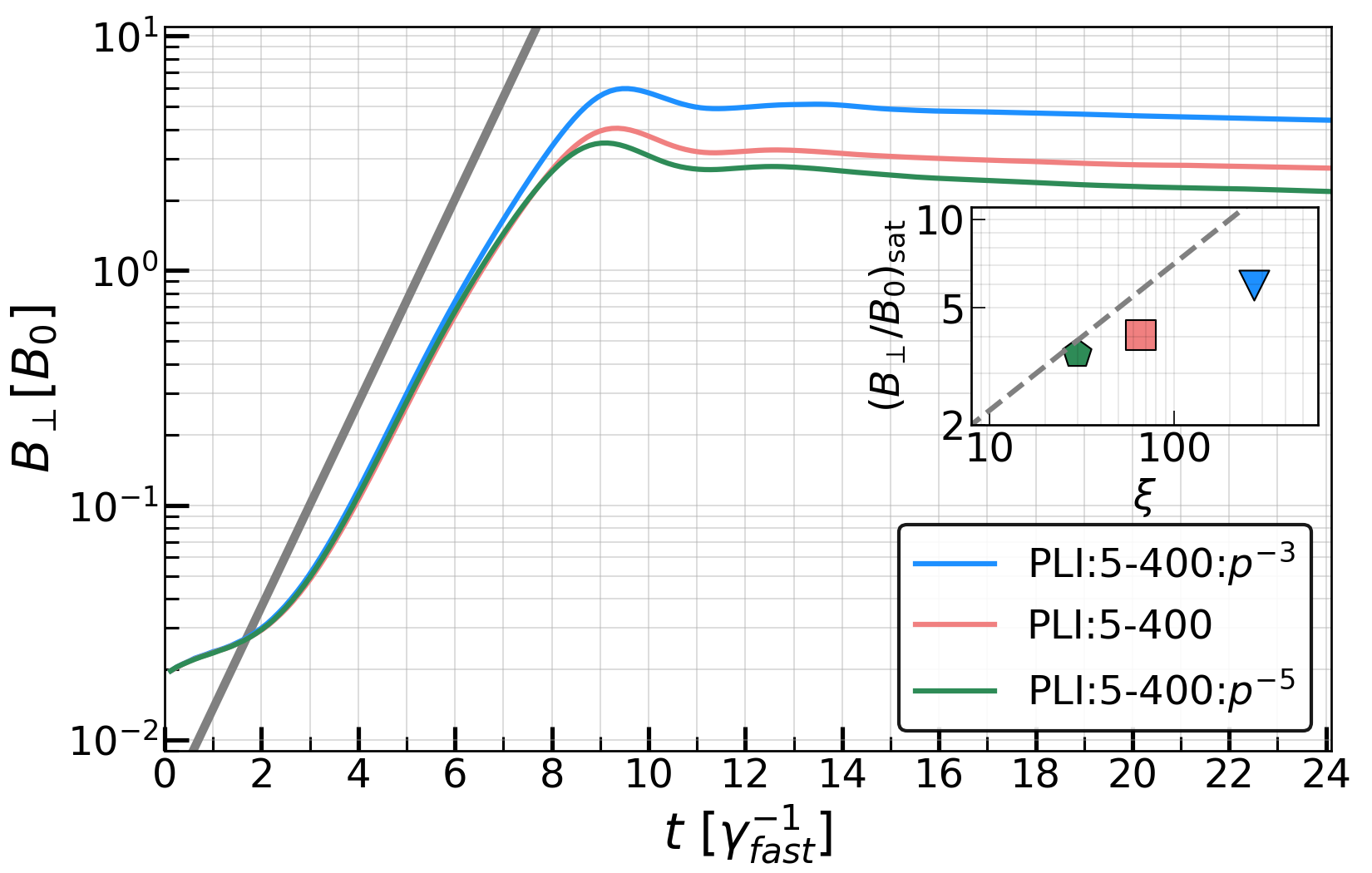}
    \caption{Evolution of transverse magnetic field $B_\perp$ for the runs PLI:5-400:$p^{-3}$, PLI:5-400($p^{-4}$), and PLI:5-400:$p^{-5}$.
    Inset: Saturated magnetic field $\left( B_{\perp}/B_0 \right)_{\text{sat}}$ versus anisotropy parameter $\xi$ for these runs (green pentagon for $p^{-5}$, pink square $p^{-4}$, blue nabla for $p^{-3}$). 
    The grey dashed line represents Equation \ref{eq:xi_vs_saturation}.}
    \label{fig:magfield_pressure_diffPLindex}
\end{figure}

\begin{table}
  \begin{threeparttable}
    \caption{Runs with different power-law indices}
    \centering
     \begin{tabularx}{\linewidth}{llcccc}
        \toprule
        \hline
        Sl. & Runs & $v_d$ & $p_d$ & $\xi$ & $k_{\text{fast}}d_i$ \\
        No. & & $[c]$ & $[m_ic]$ & & $(\times 10^{-2})$ \\ 
        \midrule
        \hline
        A1 & PLI:5-400:$p^{-5}$   & 0.3 & 4.3  & 30  & 6.97 \\
        A2 & PLI:5-400:$p^{-3}$   & 0.3 & 39.2 & 271 & 6.91 \\
        \bottomrule
     \end{tabularx} \label{tab:additional_sims}
    \begin{tablenotes}
      \small
      \item \textbf{Note:} The labels of the runs and the parameters are the same as those given in Table \ref{tab:sims_data} and Table \ref{tab:grid_field_particle}.
      Additionally, in the labels, the CR distribution $f'_{\rm cr}(p) \propto p^{-5}$ and $p^{-3}$ are mentioned.
      Movies for these runs are provided (\href{https://www.youtube.com/playlist?list=PLsIECy7LXbWI22xY_-57BXBjiLnbcP1mM}{Click here}).
    \end{tablenotes}
  \end{threeparttable}
\end{table}

DSA in a strong shock produces a power-law distribution of CRs with $f'_{\text{cr}} (p) \propto p^{-4}$.
However, the power-law index can differ when the shock is not highly supersonic or is modified by CR transport. 
We investigate how a change in the power-law index affects NRSI.
We consider isotropic CRs with $g'(p) \propto p^{-5}$ and $p^{-3}$ (see Equation \ref{eq:distribution_functions}). 
We perform the runs PLI:5-400:$p^{-5}$ and PLI:5-400:$p^{-3}$ (A1 and A2 in Table \ref{tab:additional_sims}) with these CR distributions.
We also consider PLI:5-400 (I4 in Table \ref{tab:sims_data}), for which $f'_{\rm cr} (p) \propto p^{-4}$.

We begin with the linear regime.
For a fixed $n_{\rm cr}/n_0$, $k_{\rm fast}$ and $\gamma_{\rm fast}$ (Equation \ref{eq:k_fast_fluid}) depend on drift speed $v_d$.
For relativistic CRs, $v_d$ depends only on $\mu'_{\min}$ and $v_b$ (Equation \ref{eq:drift_vel_approx}).
Thus, the linear regime is independent of the power-law indices (cf. $k_{\rm fast}$ of A1, A2, and I4). 
Figure \ref{fig:magfield_pressure_diffPLindex} shows the evolution of transverse magnetic fields $B_\perp = \sqrt{B_y^2+B_z^2}$, normalized to the initial magnetic field $B_0$, for these runs.
The exponential growths for PLI:5-400:$p^{-5}$ and PLI:5-400:$p^{-3}$ are similar to PLI:5-400.
The other diagnostics are similar to Figure \ref{fig:MagFld_Snaps_pcr_5}.

We test the saturation prediction in Equation \ref{eq:xi_vs_saturation} for these runs.
We evaluate CR momentum drift $p_d$ (Equations \ref{eq:mean_momentum}) and anisotropy parameter $\xi$ (Equation \ref{eq:xi_gupta_2021}).
The inset of Figure \ref{fig:magfield_pressure_diffPLindex} shows $(B_{\perp}/B_0)_{\rm sat}$ versus $\xi$. 
As the distribution becomes steeper, $\xi$ decreases, resulting in a smaller $(B_{\perp}/B_0)_{\rm sat}$.
However, $(B_{\perp}/B_0)_{\rm sat}$ for $f'_{\rm cr}(p) \propto p^{-3}$ deviates more from the prediction than $f'_{\rm cr}(p) \propto p^{-5}$.
This is due to numerous higher-energy CRs in PLI:5-400:$p^{-3}$, which take longer to isotropize and therefore do not contribute effectively to the saturated field.
In other words, the quantity $p'_{\rm eff}$ (Equation \ref{eq:p_cut}) is smaller for $f'_{\rm cr} (p) \propto p^{-5}$ and larger for $f'_{\rm cr}(p) \propto p^{-3}$, compared to $f'_{\rm cr}(p) \propto p^{-4}$.

\section{NRSI by nonrelativistic CRs} \label{sec:nonrel_crs}

\begin{table}
  \begin{threeparttable}
    \caption{Runs with nonrelativistic CRs}
    \centering
     \begin{tabularx}{\linewidth}{llccccc}
        \toprule
        \hline
        Sl. & Runs & $n_{\rm cr}$ & $v_b$ & $v_d$ & $\xi_{\rm eff}$ & $k_{\text{fast}}d_i$ \\
        No. & & $[n_0]$ & $[c]$ & $[c]$ & & $(\times 10^{-2})$ \\ 
        \midrule
        \hline
        N1 & MEI:0.25   & $10^{-2}$ & 0.4 & 0.4 & 44  & 22.9 \\
        N2 & MEI:0.4   & $2 \times 10^{-2}$ & 0.2 & 0.2 & 21 & 28.7 \\
        N3 & PLI:0.25-25   & $10^{-2}$ & 0.4 & 0.4 & 49 & 28.1 \\
        N4 & PLI:0.4-40   & $2 \times 10^{-2}$ & 0.2 & 0.2 & 27 & 26.0 \\
        N5 & PLC:0.4-40   & $2 \times 10^{-2}$ & 0   & 0.3 & 87 & 52.6 \\
        \hline
        R1 & MEI:8   & $8 \times 10^{-3}$ & 0.4   & 0.3 & 20 & 5.4 \\
        R2 & PLI:5-30   & $10^{-2}$ & 0.4   & 0.3 & 32 & 6.8 \\
        \bottomrule
     \end{tabularx} \label{tab:nonrel_sims}
    \begin{tablenotes}
      \small
      \item \textbf{Note:} The labels of the runs and the parameters are similar to Table \ref{tab:sims_data} and Table \ref{tab:grid_field_particle}.
      The runs N1-N5 consist of nonrelativistic CRs.
      Additionally, the runs R1 and R2 consist of relativistic CRs with roughly similar $\xi_{\rm eff}$ as N2 and N4, respectively. 
      Some of the simulation parameters for the runs N1-N5 with nonrelativistic CRs are different, which are as follows: 
      $\Delta x = 0.02 d_e$, $\Delta t = 0.009 \omega_{pe}^{-1}$, $v_{A0} = 0.006c$, and $v_{{\rm th},i} = 0.003c$.
      
    \end{tablenotes}
  \end{threeparttable}
\end{table}

\begin{figure}[!htbp]
    
    \centering
    \includegraphics[width=0.47\textwidth]{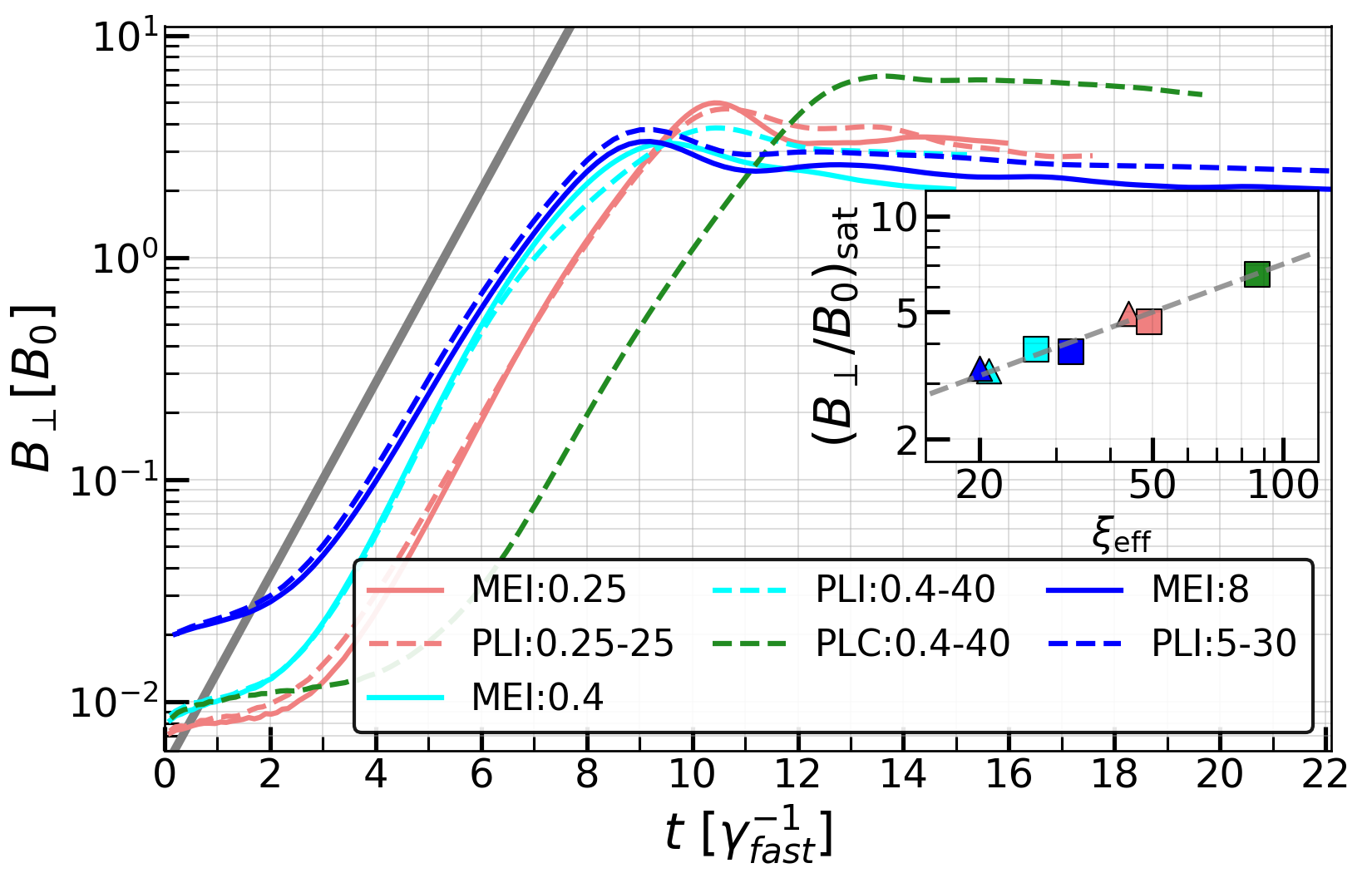}
    \caption{Evolution of transverse magnetic fields $B_\perp$ for the runs in Table \ref{tab:nonrel_sims} with nonrelativistic CRs.
    The results are similar to the runs with relativistic CRs (see Section \ref{sec:results}); we include two of these runs for comparison.  
    Inset: Saturated magnetic field $\left( B_{\perp}/B_0 \right)_{\text{sat}}$ versus effective anisotropy parameter $\xi_{\rm eff}$ for these runs (triangles for mono-energetic CRs, squares for power-law CRs; the colors are similar to those in the legend). 
    The grey dashed line represents Equation \ref{eq:saturation_vs_xi_eff}.}
    \label{fig:magfield_nonrel}
\end{figure}

We perform simulations to study NRSI driven by nonrelativistic CRs and to compare the results with the relativistic CRs.
Table \ref{tab:nonrel_sims} lists the runs.
Figure \ref{fig:magfield_nonrel} shows the evolution of the transverse magnetic fields $B_{\perp} = \sqrt{B_y^2+B_z^2}$ and the saturated fields $(B_{\perp}/B_0)_{\rm sat}$.
For all the runs, the saturated fields scale with the effective anisotropy parameter $\xi_{\rm eff}$ as Equation \ref{eq:saturation_vs_xi_eff}.
The results are similar to the runs with relativistic CRs (see Section \ref{sec:results}).

We consider the runs with mono-energetic nonrelativistic CRs (N1 and N2 in Table \ref{tab:nonrel_sims}).
Figure \ref{fig:magfield_nonrel} shows that $(B_{\perp}/B_0)_{\rm sat}$ effectively scales with the anisotropy parameter.
Notably, this is independent of whether $v_b>v'_{\rm cr}$ (e.g. N1) or $v_b<v'_{\rm cr}$ (e.g. N2).

Next, we consider the power-law distributions with low-energy CRs being nonrelativistic (N3-N5 in Table \ref{tab:nonrel_sims}).
We evaluate the effective anisotropy parameters $\xi_{\rm eff}$ using Equation \ref{eq:xi_effective}, with $p'_{\rm eff} \approx 9.3 p'_{\min}$ (Equation \ref{eq:p_cut}), similar to the relativistic CRs. 
Figure \ref{fig:magfield_nonrel} shows that $(B_{\perp}/B_0)_{\rm sat}$ effectively scales with $\xi_{\rm eff}$.
For comparison, we include two runs with relativistic CRs having similar $\xi_{\rm eff}$.
Thus, nonrelativistic CRs can amplify magnetic fields via NRSI, as do relativistic CRs.
Our results are valid for both.

\end{document}